\documentclass[11pt]{article}
\usepackage[
    top=1in,
    bottom=1in,
    left=0.8in,
    right=0.8in
]{geometry}
\usepackage{graphicx} 
\usepackage{xcolor}
\usepackage{titling}
\usepackage{multirow}
\usepackage{float}
\usepackage[skip=5pt]{caption} 
\usepackage{enumitem}
\usepackage{url}
\usepackage[
  colorlinks=true,
  linkcolor=blue,
  citecolor=blue,
  urlcolor=blue
]{hyperref}

\usepackage{lineno}

\usepackage{cite} 

\makeatletter
\let\cite@orig\cite
\renewcommand{\cite}[1]{\textsuperscript{\cite@orig{#1}}}
\renewcommand\@biblabel[1]{#1.}
\makeatother

\usepackage{soul}

\usepackage{lmodern}
\usepackage{amsmath}

\title{A flexible language model-assisted \\electronic design automation framework}

\author{
    Cristian Sestito\textsuperscript{1,*},
    Panagiota Kontou\textsuperscript{1},
    Pratibha Verma\textsuperscript{1,2}, \\
    Atish Dixit\textsuperscript{1},
    Alexandros D. Keros\textsuperscript{1},\\
    Michael O'Boyle\textsuperscript{3},
    Christos-Savvas Bouganis\textsuperscript{4},
    Themis Prodromakis\textsuperscript{1}
}

\date{
    \small
    \textsuperscript{1}Centre for Electronics Frontiers, Institute for Integrated Micro and Nano Systems, \\ School of Engineering, The University of Edinburgh, UK \\
    \textsuperscript{2}Department of Electrical Engineering, Indian Institute of Technology Indore, India \\
    \textsuperscript{3}School of Informatics, The University of Edinburgh, UK \\
    \textsuperscript{4}Department of Electrical and Electronic Engineering, Imperial College London, UK \\[0.2in]
    \textsuperscript{*}Correspondence: csestito@ed.ac.uk \\
}

\begin{document}

\maketitle


\begin{abstract}
Large language models (LLMs) are transforming electronic design automation (EDA) by enhancing design stages such as schematic design, simulation, netlist synthesis, and place-and-route. Existing methods primarily focus these optimisations within isolated open-source EDA tools and often lack the flexibility to handle multiple domains, such as analogue, digital, and radio-frequency design. In contrast, modern systems require to interface with commercial EDA environments, adhere to tool-specific operation rules, and incorporate feedback from design outcomes while supporting diverse design flows. We propose a versatile framework that uses LLMs to generate files compatible with commercial EDA tools and optimise designs using power-performance-area reports. This is accomplished by guiding the LLMs with tool constraints and feedback from design outputs to meet tool requirements and user specifications. The framework is used to evaluate case studies on operational transconductance amplifiers, microstrip patch antennas, and FPGA circuits.
\end{abstract}

\section*{Introduction}

Electronic design automation (EDA) is crucial for designing circuits, such as analogue amplifiers, digital architectures, and radio-frequency (RF) modules\cite{Ozalas_2025,Dall_Ora_2025,Li_2022}. Indeed, EDA tools provide automation support from the early schematic stages to the physical implementation, as well as assistance with design simulation and analysis. However, engineers often require multiple iterations to meet design specifications, involving careful analyses of log files and power-performance-area (PPA) reports to identify errors or improve implementations.

Generative AI, particularly large language models (LLMs), is transforming this iterative design process from a purely human-EDA interaction to a collaborative human-LLM-EDA approach\cite{Pan_2025,Gao_2025,Mangalagiri_2024,Grus_2024,Sharma_2025,Zadeh_2025,Guven_2025,Gielen_2025,Sapatnekar_2023,Xu_2019,Jin_2025,Molnar_2025,Mehradfar_2025,Qin_2025,Kimia_2025,Wu_2024,Qiu_Siyu_2024,Blocklove_2023}, with expectations of a tenfold increase in productivity within the electronics industry\cite{Brown_2023}. LLMs effectively assist in various design stages, such as register-transfer-level code generation\cite{Blocklove_2025,Nakkab_2024,Tang_2025}, testbench generation\cite{Qiu_2024, Chen_2025}, Spice netlist generation\cite{Bhandari_2025}, transistor sizing\cite{Ghosh_2025,Chang_2025}, floorplanning and placement\cite{Liu_2025}, static timing analysis\cite{Wenhao_2025}, bonding pad insertion\cite{Zhu_2025}, design verification\cite{Qayyum_2024,Wang_Xi_2024,Kang_2025}, design code fix\cite{Wang_2025,Qayyum_Khushboo_2025,Xu_2025}, end-to-end high level synthesis\cite{Fu_2023}. 

These advancements have primarily focused on integrating LLMs with open-source EDA tools tailored to specific design flows. However, the increasing complexity of modern electronic systems requires comprehensive support across multiple domains, including analogue, digital, and RF design. Addressing this challenge requires flexible frameworks that can seamlessly interface with commercial EDA tools and manage diverse design requirements. In this context, the effective use of LLMs depends on carefully crafted prompts, which guide the models to generate outputs that comply with tool-specific syntax and operational rules, as well as meeting user-defined design specifications. Furthermore, effective automation requires that LLMs receive feedback from EDA tool outputs, such as simulation results and PPA metrics. This enables the models to assess design outcomes and offer informed optimisation recommendations, ensuring reliable and practical integration into real-world design workflows.

In this article we propose LaMDA: a flexible framwework for \textbf{LA}nguage \textbf{M}odel-assisted electronic \textbf{D}esign \textbf{A}utomation. LaMDA leverages commercial LLMs across three common circuit design flows: analogue design, RF design, and field-programmable gate array (FPGA) design. These flows were selected because they represent widely used domains in modern electronic systems and capture the key challenges encountered in practical design workflows. The supported LLMs generate EDA-compliant source files, such as analogue netlists, RF netlists, and Verilog hardware description language (HDL) scripts. These files are used by commercial EDA tools, including Cadence Virtuoso Spectre, Keysight Advanced Design System (ADS), and AMD Vivado, which are interfaced through command scripts for batch-mode operation. Additionally, LaMDA processes PPA reports and log files and can supply LLMs with the critical information required for design optimisation. Beyond acting as a flexible framework that integrates with diverse EDA tools, LaMDA serves as an interactive, knowledge-driven assistant capable of providing informed guidance. 

LaMDA provides two operational modes. In the first mode, the framework assists engineers with custom design and optimisation problems by leveraging LLMs to analyse design data, suggest improvements, and explore alternative solutions. In the second mode, LaMDA benchmarks large datasets to evaluate the LLMs' ability to generate contextually relevant outputs across a wide range of design scenarios. To demonstrate these capabilities, we evaluate LaMDA across three case studies: (a) the design, simulation, and analysis of operational transconductance amplifiers (OTA) and inverters; (b) the design, simulation, and analysis of microstrip patch antennas; and (c) the benchmarking of digital circuits described at the Verilog HDL level. Together, these case studies highlight both the framework’s ability to support practical, domain-specific engineering tasks and its capacity to assess LLM performance across diverse circuit types.

\section*{The LaMDA framework}

\begin{figure}[t]
\includegraphics[width=\textwidth]{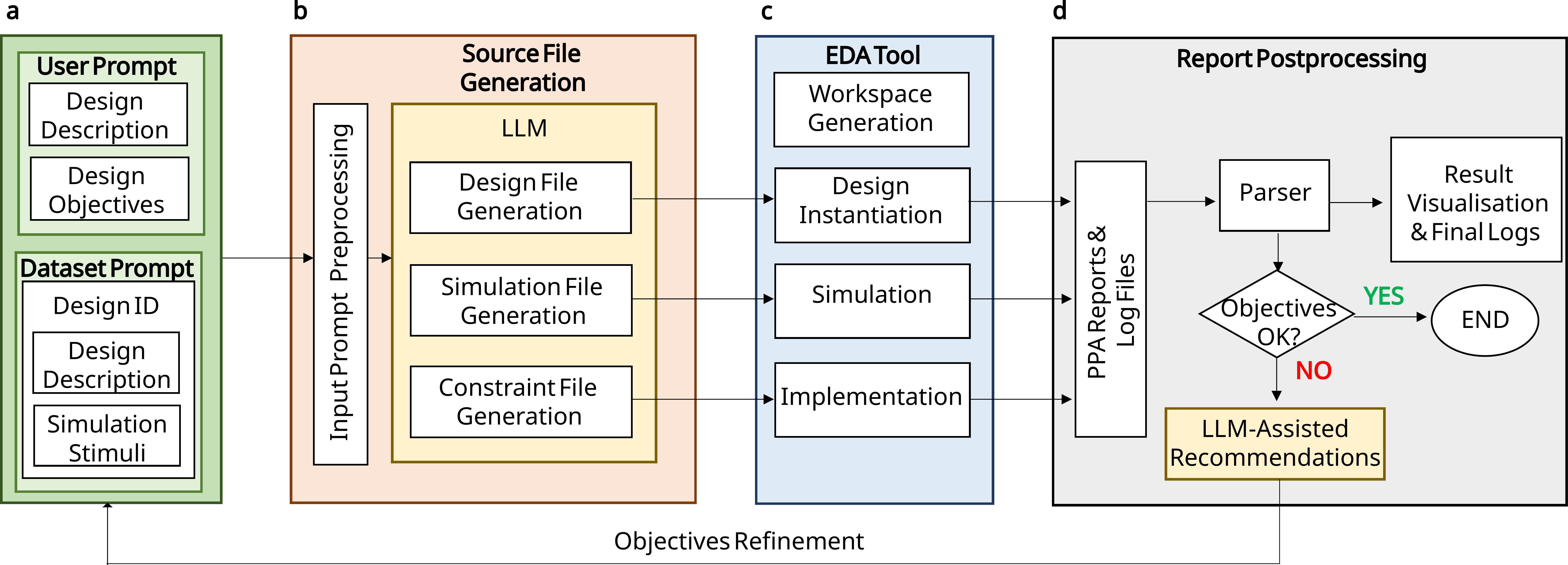}
\centering
\caption{
\textbf{A block diagram of the LaMDA framework}. 
\textbf{a,} The framework processes either a custom user prompt or a predefined prompt from a dataset. The custom user prompt includes design description and design objectives, such as area, delay, power constraints. The predefined prompt from a dataset may also include additional information, such as simulation stimuli.
\textbf{b,} The Source File Generation module includes an LLM that provides source files for the EDA tool, such as analogue and RF netlists or Verilog scripts. The module also preprocesses input prompts to record the output token count and time corresponding to the LLM-generated response.
\textbf{c,} The EDA tool is provided with source files from the LLM, generates the workspace, instantiates the design, runs simulations, and executes the design implementation. 
\textbf{d,} The PPA reports and log files are supplied to the Report Postprocessing module. Key log messages and information such as occupied area and power dissipation are extracted through a parser. The parsed information is logged and can be visualised for further data analysis. The framework supports two execution paths: termination of the process or continuation through additional optimisation cycles with LLM-assisted recommendations. In either case, the objectives specified in the input prompts are evaluated.
}
\label{Figure_1}
\end{figure}

The operational workflow of LaMDA framework is illustrated in Fig. \ref{Figure_1}. Custom design prompts consist of a user prompt (Fig. \ref{Figure_1}a), which specifies the circuit design and constraints, and a system prompt, which provides the tool-specific information needed to guide the LLM (Supplementary Figs. \ref{Figure_AS1}a, \ref{Figure_RS0}a, \ref{Figure_FS1}a). Within the analogue design domain, we leverage the first operational mode of LaMDA. The user can request LaMDA framework to assist in designing an OTA that meets a target gain and achieves a stable phase margin (Fig. \ref{Figure_2}a). To realise this task, the LLM interfaces with Cadence Virtuoso Spectre. The LLM is provided with a system prompt that specifies (a) Spectre-compliant netlist syntax rules, (b) allowable transistor sizing ranges, and (c) definitions of the relevant performance metrics (Supplementary Fig. \ref{Figure_AS1}a). Building on this approach, LaMDA framework can also be applied to RF design tasks. In this case, the user can be assisted in implementing antennas that operate at a target frequency while satisfying a specified reflection coefficient performance (Supplementary Fig. \ref{Figure_RS0}a). For RF circuit design, Keysight ADS is supported. The system prompt includes the necessary Keysight ADS information, specifically: (a) the available components and libraries, (b) representative netlist examples, and (c) ADS Python interface specifications (Supplementary Fig. \ref{Figure_RS0}a). The framework’s second operational mode is exploited as a research platform for benchmarking LLMs across large datasets comprising tens or even hundreds of design prompts. These prompts can include design descriptions, specifications (e.g., look-up table (LUT) count, maximum delay, clock frequency), and simulation stimuli in Verilog HDL (Fig. \ref{Figure_1}a and Supplementary Fig. \ref{Figure_FS1}a). When digital design using FPGAs is considered, the user can evaluate datasets that include finite state machines (FSMs) (Fig. \ref{Figure_4}a) to study the capacity of LLMs to provide correct Verilog scripts.

To translate these prompts into simulation- or implementation-ready source files for supported EDA tools, the user can access a library of software modules that facilitate interaction with commercial LLMs. The selected LLM is then responsible for generating the required design files, including design, simulation, and hardware constraint files, as depicted in Fig. \ref{Figure_1}b. In the analogue domain, the LLM generates a complete Spectre netlist with parameterised device dimensions and simulation directive(Supplementary Fig. \ref{Figure_AS1}b). LaMDA framework generates Keysight ADS-compliant netlists for the RF design task and provides comments and design recommendations. To further ensure correctness, a graph-based netlist visualisation is produced to verify proper component connectivity (Supplementary Fig. \ref{Figure_RS0}b). In the case of FPGA design, the LLM outputs a Verilog HDL script (Supplementary Fig. \ref{Figure_FS1}b). The LaMDA framework also includes software modules to further preprocess the input prompts, for example, extracting a Verilog HDL testbench file and hardware constraints to be supplied to the EDA tool (Supplementary Fig. \ref{Figure_FS1}b).

LaMDA includes a library of command scripts to run EDA tools in batch-mode for the instantiation, simulation, and implementation of the design (Fig. \ref{Figure_1}c). Cadence Virtuoso Spectre is driven by Python scripts that bind the LLM-generated Spectre decks to the foundry models and run DC, AC, and transient analysis (Supplementary Fig. \ref{Figure_AS1}c). Similarly, LaMDA framework includes Keysight ADS Python scripts that automatically create a workspace, library, schematic, and simulate the LLM-generated netlist (Supplementary Fig. \ref{Figure_RS0}c). In the case of FPGA design, LaMDA generates tool command language (TCL) script for the AMD Vivado tool to instantiate, simulate, synthesise, and implemented the target circuit (Supplementary Fig. \ref{Figure_FS1}c).

For each design step executed by the EDA tool, key information from the power-performance-area (PPA) report files and log files is parsed while information are logged for visualisation and further data analysis (Fig. \ref{Figure_1}d). When the analogue design is considered, the parser reads the Spectre Parameter Storage Format (PSF) outputs and log files to extract key metrics and checks whether the design specifications are met (Supplementary Fig. \ref{Figure_AS1}d). For RF design, the framework supports two optimisation strategies. In the first, a closed-loop optimisation is executed over a predefined number of iterations, where each iteration consists of simulation, result evaluation, and user-guided LLM updates that generate an updated netlist. In the second strategy, the optimisation proceeds iteratively until the simulated or implemented design satisfies the specified performance requirements, at which point the process terminates (Supplementary Fig. \ref{Figure_RS0}d). Eventually, in the case of FPGA design, reports related to resource occupation, delays, and power consumption are parsed. In addition, errors and critical warnings are also extracted to identify any issue during the design stages (Supplementary Fig. \ref{Figure_FS1}d). Output checks and error handling is detailed in Supplementary Table \ref{Table_S2}.

Overall, LaMDA proposes a flexible pipeline where LLMs assist EDA tools in terms of design artifacts generation, specification definitions, and design refinement assistance. Supplementary Table \ref{Table_S3} summarises the key roles of the LLM in each supported design flow and the degree of autonomy offered by the framework.

\section*{Analogue design flow application}

\begin{figure}[t]
\includegraphics[width=\textwidth]{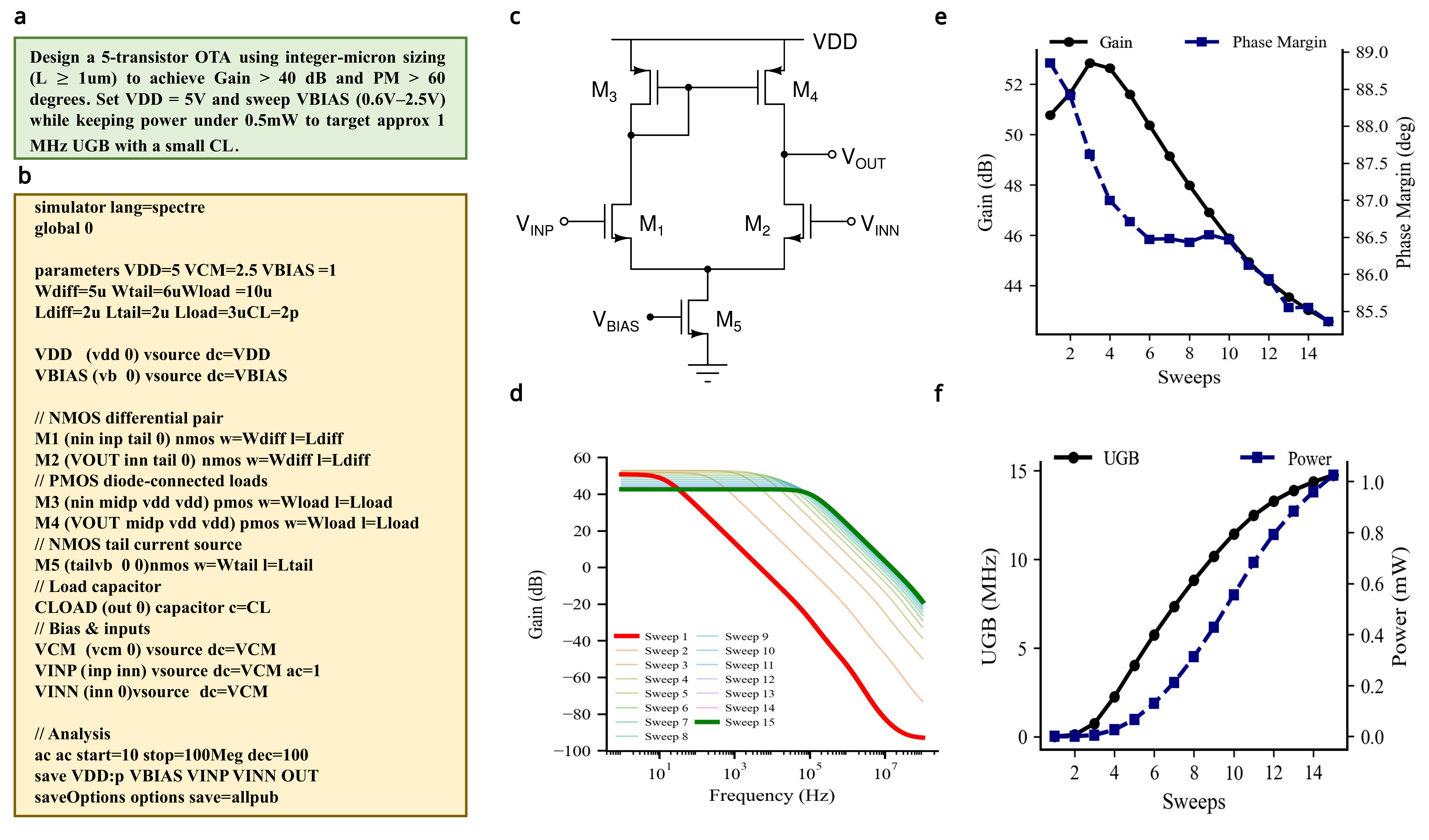}
\centering
\caption{
\textbf{Application of LaMDA in an analogue design flow example.} 
\textbf{a,} Input prompt. Design a 5--transistor OTA using integer-micron sizing (L $\geq$\,1um) to achieve Gain $>$ 40 dB and PM $>$ 60 degrees. Set VDD = 5\,V and sweep $V_{BIAS}$ (0.6V--2.5\,V) while keeping power under 0.5\,mW to target approx 1\,MHz UGB with a small CL.
\textbf{b,} Generated netlist. The netlist includes information about the device parameters ($W_{\mathrm{diff}}$, $L_{\mathrm{diff}}$, etc.) and includes necessary analysis commands and voltage sources, demonstrating the LLM's capability to generate functional SPICE code. The generated netlist is compatible with Cadence Virtuoso Spectre. LaMDA then performs 15 simulations, sweeping $V_{BIAS}$ across different values to evaluate the design and identify operating points that meet the initial objectives.
\textbf{c,} The 5-transistor OTA circuit with an NMOS differential pair (M1, M2), PMOS diode-connected loads (M3, M4), and an NMOS tail current source (M5).
\textbf{d,} OTA gain from simulations. The graph shows the reduced DC gain and increased UGB with increasing bias, illustrating the speed--gain trade-off. 
\textbf{e,} DC gain and phase margin across 15 sweeps. Gain decreases with increasing tail current, while PM remains more than $60^\circ$, confirming stability. 
\textbf{f,} UGB across 15 sweeps. UGB improves with iteration, but at the cost of steadily increasing power, with bandwidth gains gradually saturating at later sweeps. }
\label{Figure_2}
\end{figure}

For evaluation of the LaMDA framework on analogue circuit design, we considered the sizing of a 5-transistor OTA (5T OTA). This circuit was selected as the primary evaluation benchmark because it represents a canonical building block in analogue design. It requires the LaMDA framework to resolve classic design trade-offs, specifically the coupling between DC gain, unity gain bandwidth, and stability. The design target was set to a low-frequency gain $>$ 40 dB, phase margin $>$ 60°, and unity gain bandwidth (UGB) $\approx$ 1 MHz. The input prompt is shown in Fig. \ref{Figure_2}a. Guided by user prompt and the system prompt, the LLM generated a single Spectre-compatible simulation deck that combines the parametrised transistor-level OTA netlist (device sizes and bias parameters), the testbench sources and loads, and the analysis directives required to run simulations, as shown in Fig. \ref{Figure_2}b. On average, the LLM required 16 seconds to generate the netlist. Simulations performed using Cadence Spectre resulted in only a minimal overhead of approximately 1 second per sweep. Regarding the token count, the LLM response used 1200 tokens on average. The corresponding 5T OTA schematic used as the target topology is shown in Fig. \ref{Figure_2}c.

The generated Spectre deck was simulated in batch mode, and the resulting PSF outputs and log files were parsed to extract the key performance metrics. We tracked the gain, the corresponding stability margin (phase margin) and UGB across sweep points corresponding to different $V_{BIAS}$ values. 
As shown in Fig. \ref{Figure_2}d, increasing the bias point shifts the unity-gain bandwidth to higher frequency while reducing the low-frequency (DC) gain, highlighting the inherent speed–gain trade-off. The gain and phase margin requirement were satisfied across all sweeps (Fig. \ref{Figure_2}e). Fig. \ref{Figure_2}f shows the variation of UGB and power consumption across bias voltage sweep points, showing a monotonic increase in both metrics and increasing $V_{BIAS}$. Based on these extracted performance metrics, LaMDA simultaneously computes and visualises the key design trade-offs from the generated netlist.
We further validated the prompt sensitivity for OTA. We gave a less detailed prompt to LaMDA (Supplementary Fig. \ref{Figure_AS2}), and in this case the gain requirement was
satisfied only in the early iterations and was violated at higher-bias iterations, whereas the phase margin
remained approximately constant near 90° across the sweep and the power consumption increased.

In addition to the 5T OTA case study, we validated the analogue design flow considering a CMOS inverter sizing study and explicitly examining the prompt sensitivity. In the less detailed prompt setting (Supplementary Fig. \ref{Figure_AS3}), we requested a low-power inverter and a $W_P/W_N$ sweep from 2.0 to 3.0,
without specifying a concrete optimisation objective, a delay definition, or noise-margin
constraints. The extracted metrics and the returned set of design points reflected a broader range of operating conditions, leading to a scattered delay-power distribution and an unclear set of optimal designs configurations.
In contrast, in the detailed prompt reported in Supplementary Fig. \ref{Figure_AS4}, we provided a more explicit objective (power delay product (PDP) minimisation), a precise delay definition ($\max(t_{pHL}, t_{pLH})$ measured at 50\% $V_{DD}$), a constrained sizing search (same $W_P/W_N$ sweep), and functional correctness constraints via noise margins ($\mathrm{NMH} \ge 1.25~\mathrm{V}$ and $\mathrm{NML} \ge 1.25~\mathrm{V}$ extracted from the DC transfer curve). These additional specifications directly tightened the measurement extraction and filtering logic, yielding a compact and consistent trade-off curve and a clearer Pareto front over power versus worst-case delay. Both experiments demonstrate that the response of the framework is prompt-dependent: increasing prompt specificity changes (a) which metrics are measured and how they are computed, (b) which designs are considered valid, and (c) which operating point is selected as best under the stated objective.

\section*{Radio-frequency design flow application}

\begin{figure}[t]
\includegraphics[width=\textwidth]{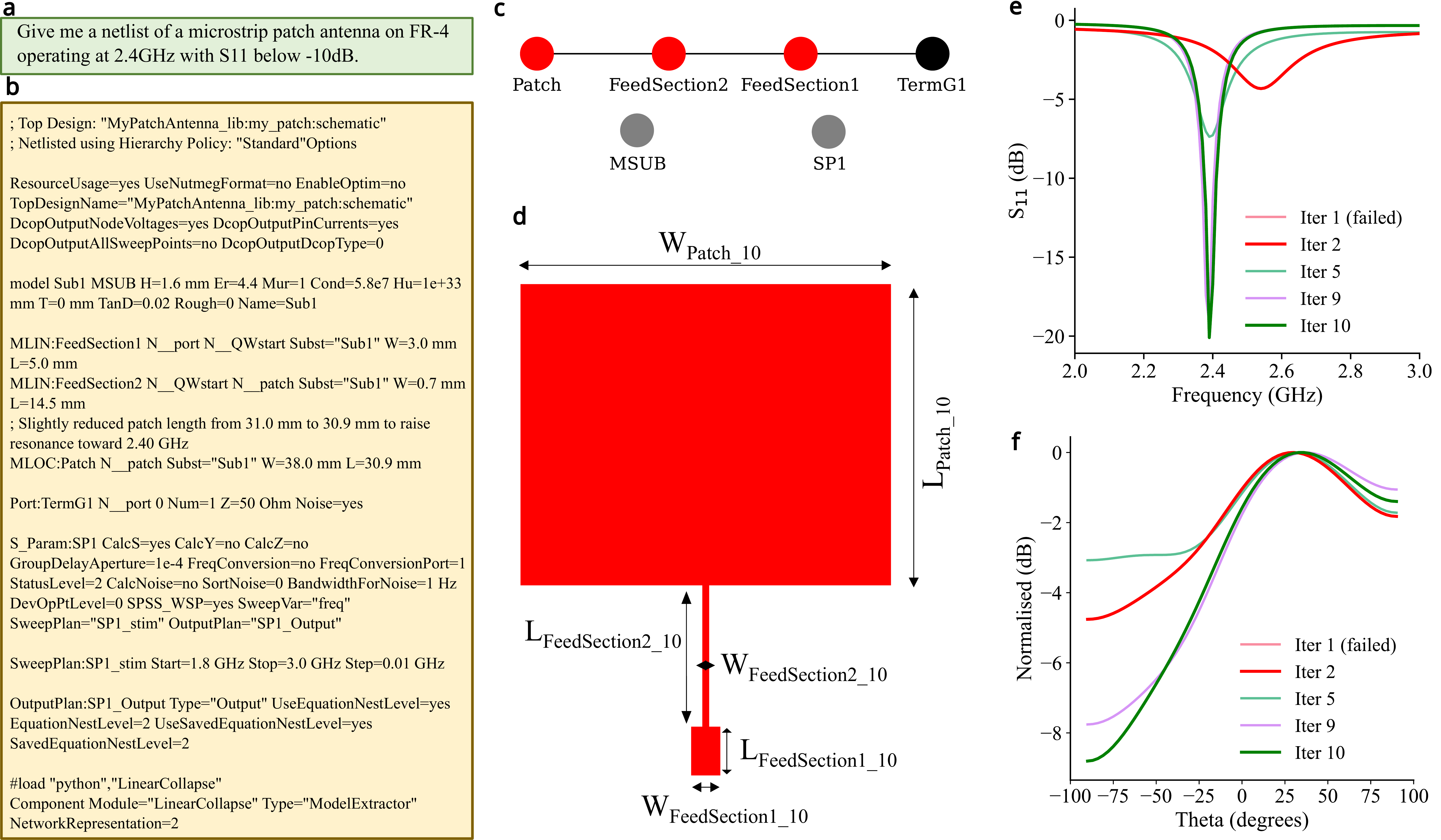}
\centering
\caption{
\textbf{Application of LaMDA in an RF design flow example.} 
\textbf{a,} Input prompt. The user specifies the design of a microstrip patch antenna on an FR-4 substrate, with an operating frequency of $f=2.4$ GHz and a target reflection coefficient $S_{11} < -10\,\mathrm{dB}$. 
\textbf{b,} Netlist generated by the LLM, which includes the antenna parameters in the form of transmission line components, the microstrip substrate parameters, and the S-parameters simulation setup. The generated netlist is compatible with Keysight ADS. Of 10 predefined iterations, iteration 1 failed due to a component's terminal specification error, therefore, iteration 2 is considered as the initial valid iteration. The target performance was achieved at iteration 9 and further improved at iteration 10. The reported netlist refers to the final iteration.
\textbf{c,} Graph-based visualisation of the netlist to ensure correct connectivity between components.
\textbf{d,} The equivalent antenna layout generated by Keysight ADS. 
\textbf{e,} Reflection coefficient from 2D S-parameters simulation. We report iterations 1, 2, 5, 9, 10. Each curve consists of 101 frequency-S$_{11}$ pairs.
\textbf{f,} Normalised far-field H-plane ($\phi=90^{\circ}$) radiation pattern of the patch antenna for the last design iteration, generated from a 3D EM simulation. The curve consists of 62 $\theta-\vert E(\theta)\vert$ pairs.
}
\label{Figure_3}
\end{figure}

To evaluate the LaMDA framework on the RF design, we considered the design of a microstrip patch antenna operating at $f=2.4$ GHz on an FR-4 substrate with a performance target of a reflection coefficient S$_{11}$ below $-10\,\mathrm{dB}$. Patch antennas were selected as an example because of their extensive use in real-world applications and the availability of well-documented performance benchmarks. The input prompt is shown in Fig. \ref{Figure_3}a. The LLM generated the circuit netlist which includes the antenna parameters in the form of transmission line components, the microstrip substrate parameters, and the S-parameters simulation setup. The generated netlist is compatible with Keysight ADS (Fig. \ref{Figure_3}b). LaMDA framework also generates a graph-based visualisation of the netlist to validate correct connectivity between components, as shown in Fig. \ref{Figure_3}c. Each node corresponds to a microstrip component of the antenna, the termination port, the substrate model, or the S-parameter simulation block. The workspace, library and schematic were created, and the LLM-generated netlist was simulated in Keysight ADS. We were interested in the reflection coefficient results, which were visualised and inspected (Fig. \ref{Figure_3}e). A total of 10 human–LLM interaction iterations (prompt, response, simulation, feedback) were predetermined and executed. Iteration 1 failed due to a runtime error related to a component's terminal specification. Therefore, iteration 2 is considered the initial valid result. The results of all iterations are presented in Supplementary Fig. \ref{Figure_RS1} and Fig. \ref{Figure_RS2}.

The antenna consists of a radiating patch and a feedline. While the patch dimensions primarily determine the resonant frequency and are therefore only minimally adjusted across iterations, the feedline type (e.g., single or double feed) and length are more frequently modified in the LLM-generated netlists to improve impedance matching and achieve the desired reflection coefficient. To illustrate this, the antenna layout was generated using Keysight ADS and is shown in Fig. \ref{Figure_3}d for the final iteration. Compared to the initial single-feed design (Supplementary Fig. \ref{Figure_RS1}), the subsequent and final designs employ a double-feed configuration proposed and implemented by the LLM. An Electromagnetic (EM) simulation was then performed to obtain the radiation pattern of the antenna (Fig. \ref{Figure_3}f). In the final ($10^{th}$) iteration, the main lobe is directed at $\theta \approx 40^{\circ}$, indicating off-broadside radiation. The maximum directivity is 6.2 dBi, while the corresponding maximum gain is 2.6 dBi, with a radiation efficiency of approximately 43\%, which reflects the additional losses captured in the EM simulation.

The framework successfully achieved the requested target performance of a reflection coefficient, S$_{11}$, below -10 dB. The required performance was achieved at iteration 9 ($S_{11} = -11.3\,\mathrm{dB}$) and further refined at iteration 10 ($S_{11} = -16.7\,\mathrm{dB}$). However, other performance metrics could be further enhanced. If desired, additional performance objectives, such as increased directivity or gain, can be specified in the prompt and incorporated into the design process to further tailor the antenna’s characteristics.

We prompted the LLM via a chat-based interface. On average, the LLM required 21 seconds to generate the netlist. The operations performed by Keysight ADS introduced a very limited overhead, approximately 3.4 seconds on average. In terms of token count, the input prompt used approximately 26000 tokens, which gradually increased  throughout the iterations, and the LLM response consisted of 1500 tokens on average. The large number of input tokens is explained by the presence of the following: (a) the system prompt, which contains Keysight ADS documentation, (b) the entire conversation history, including all previous iteration netlists, and (c) the current user message. Although the token count is still relatively large, we performed significant compression as described in the Methods.

\section*{FPGA design flow application}

\begin{figure}[t]
\includegraphics[width=\textwidth]{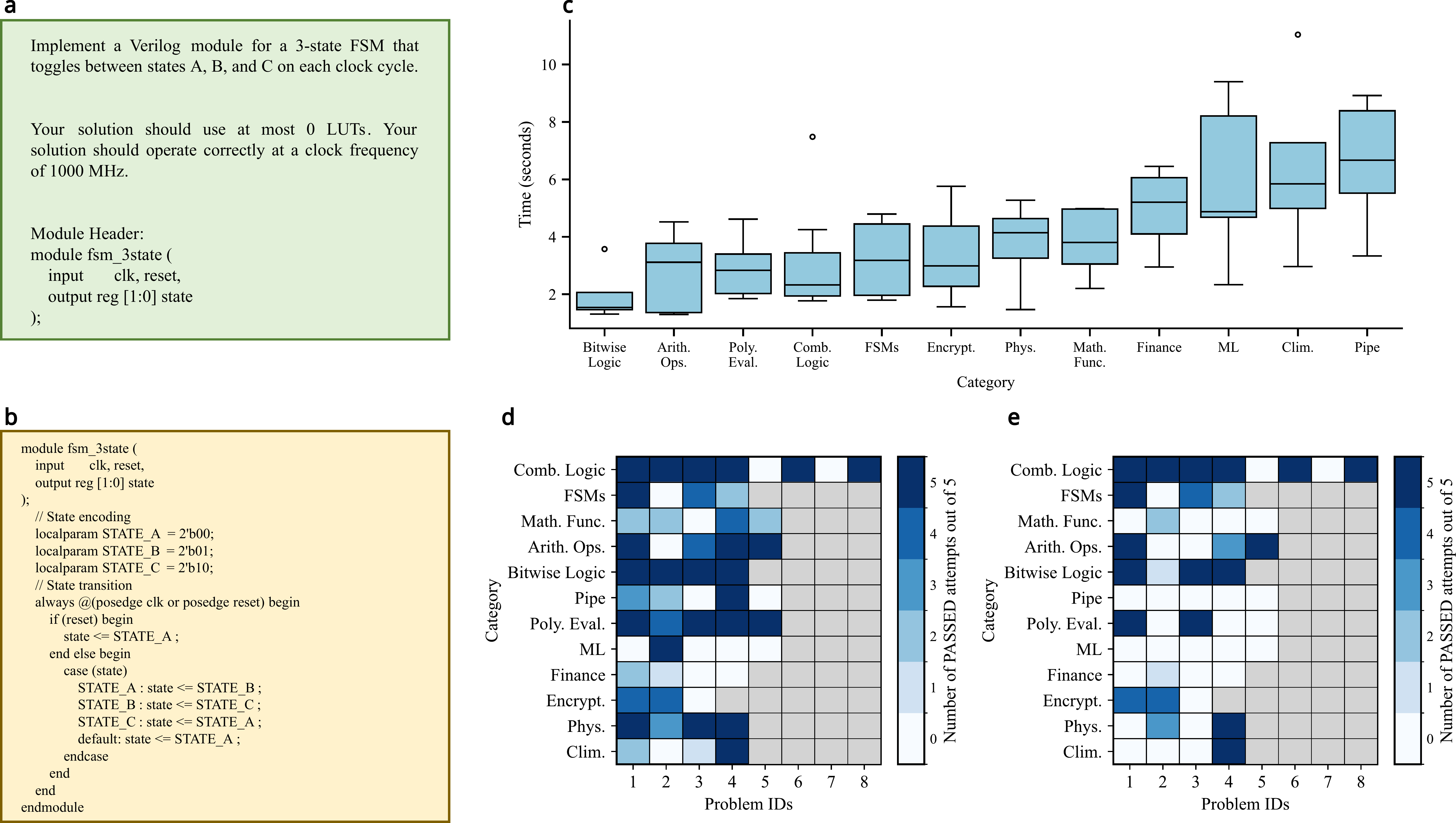}
\centering
\caption{
\textbf{Application of LaMDA in an FPGA design flow example.} 
\textbf{a,} Input prompt. The user requires a 3-state FSM, operating at a frequency of 1000 MHz and being implemented without LUTs. The module header, reporting input and output ports, is also provided. 
\textbf{b,} Verilog script generated by the LLM. 
\textbf{c,} LLM response time in seconds. The LaMDA framework runs the full ResBench dataset\cite{Guo_2025} that consists of 56 problems across 12 categories. The 3-state FSM in \textbf{a,b} is one of the problems. We run each problem 5 times to take into account the LLM randomness. On average, the LLM response time spans from 2 seconds (category: bitwise and logical operations) to 7 seconds (category: pipelining). 
\textbf{d,} Pass rate related to the implementation step on FPGA. For each category-problem ID pair, we report the number of successful implementation runs out of 5. Different categories contain different numbers of problems, therefore, grey cells are used to pad the table for visual consistency.
\textbf{e,} Pass rate related to the LUT objective. For each category-problem ID pair, we report the number of successful runs, out of 5, where the LUT objective is met.
}
\label{Figure_4}
\end{figure}

To evaluate the LaMDA framework on the FPGA design flow, we considered the ResBench dataset \cite{Guo_2025}. It consists of 56 problems across 12 categories (Supplementary Table \ref{Table_S1}). For each problem, LaMDA received a prompt template consisting of a short design description, design specifications including LUT count and clock frequency/maximum delay, and the module header with input and output ports. Fig. \ref{Figure_4}a shows the input prompt of one of the problems, which relates to a 3-state finite state machine (FSM). We specified as objectives a clock frequency of 1000 MHz and a LUT count = 0. The LLM generated the Verilog HDL design file that describes the circuit function from a behavioural point of view (Fig. \ref{Figure_4}b). Once the AMD Vivado EDA tool performed the design instantiation, simulation, synthesis, and implementation, we collected relevant information from reports and log files for further data analysis. We repeated this process for all 56 problems and ran each problem 5 times to account for the randomness of the LLM. 

We first visualised information related to LLM response time (Fig. \ref{Figure_4}c), total execution time (Supplementary Fig. \ref{Figure_FS2}), and LLM token count (Supplementary Fig. \ref{Figure_FS3}). On average, the LLM required between 2 and 7 seconds to generate Verilog HDL scripts. When the AMD Vivado time was also considered, LaMDA required between 88 and 144 seconds per problem. In terms of token count, the LLM used between 69 to 340 tokens on average. We eventually visualised the pass rate for each problem over 5 runs, and related to each design step: logic simulation (Supplementary Fig. \ref{Figure_FS4}a), design synthesis (Supplementary Fig. \ref{Figure_FS4}b), design implementation (Fig. \ref{Figure_4}d and Supplementary Fig. \ref{Figure_FS4}c), LUT objective (Fig. \ref{Figure_4}e and Supplementary Fig. \ref{Figure_FS4}d), clock frequency/maximum delay objective (Supplementary Fig. \ref{Figure_FS4}e). By examining the pass rate patterns, we observed that the LLM performance is promising until the implementation stage (with 73\% of the problem passing at least one run out of the 5 attempts). In contrast, the LLM performance degrades when the input objectives are taken into account: the 45\% of the problems satisfied the LUT objective in at least one run out of the 5 attempts, while only the 13\% of the problems satisfied the clock frequency/maximum delay objective in at least one run out of the 5 attempts. Overall, we noticed that the LLM-generated Verilog designs successfully achieve the timing requirements when a precise clock constraint is specificed through the generated artifact. Indeed, Supplementary Fig. \ref{Figure_FS4}e shows that the Finite State Machines (IDs 9, 11-12) and Pipelining (ID 30) categories met the target.

\section*{Large language models evaluation}

\begin{figure}[t]
\includegraphics[width=\textwidth]{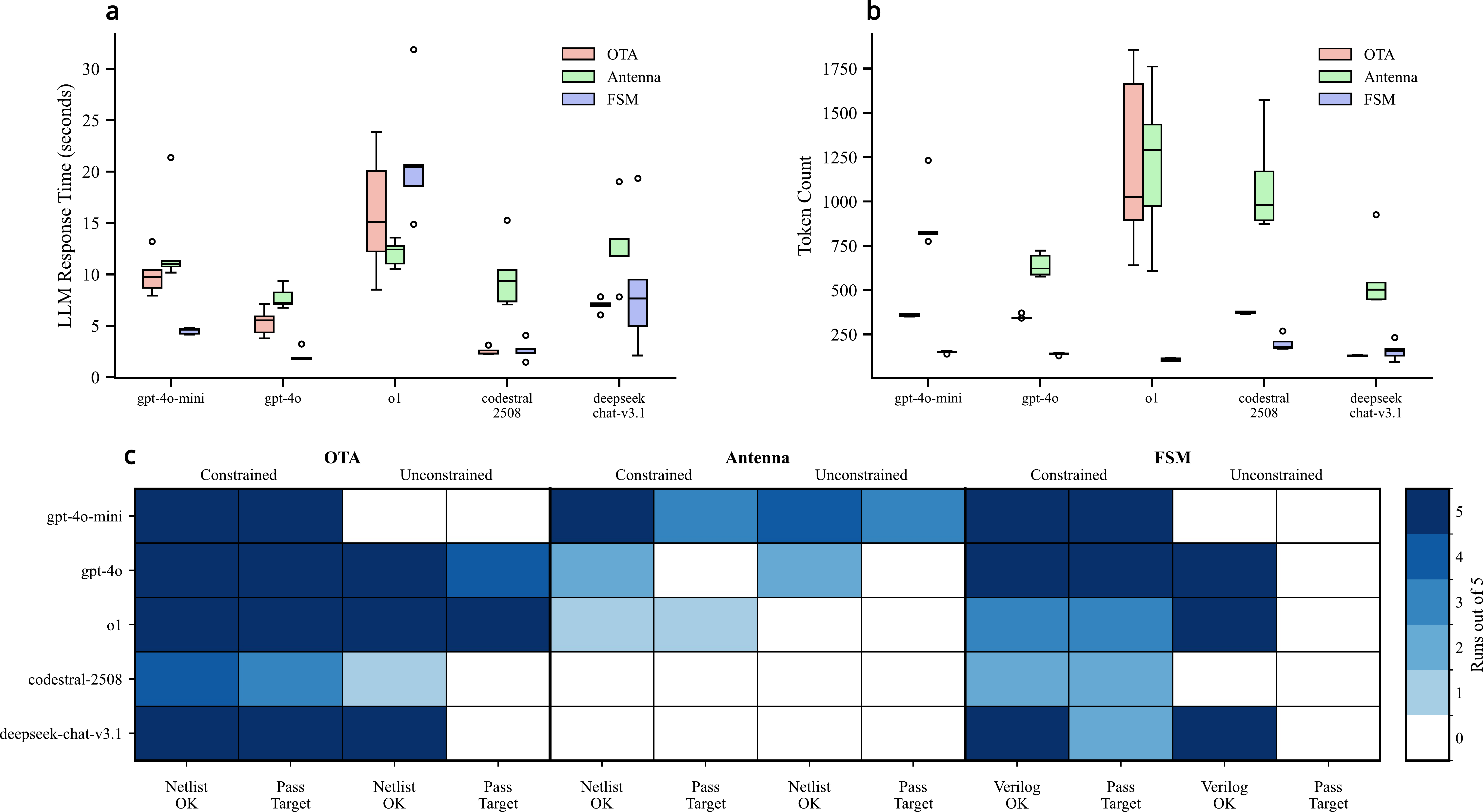}
\centering
\caption{
\textbf{Comparative analysis across different case studies (OTA, RF antenna, and FSM design) and LLM models (OpenAI GPT-4o-mini, GPT-4o, o1; Codestral 2508; DeepSeek Chat v3.1).}
\textbf{a,} LLM response time. 
\textbf{b,} LLM token count associated to the output messages. In all cases, we reported statistical results across 5 runs per model.
\textbf{c,} Success rate. Each design was assessed under two scenarios: with and without specific constraints in the user prompt. In addition, we report the scenarios where, even if the constraint is not met, the generated netlist is compliant with the related EDA tool.
}
\label{Figure_5}
\end{figure}

Figure \ref{Figure_5}a compares the LLM response time for the OTA, antenna and FSM example across five selected LLM models, including OpenAI GPT-4o, GPT-4o-mini, o1; Codestral 2508; and DeepSeek Chat v3.1.
In the OTA example, o1 exhibited the highest average response time (16 seconds) and variability, followed by GPT-4o-mini (10 seconds), while DeepSeek Chat v3.1 averaged 7 seconds, GPT-4o 5 seconds and Codestral 2508 3 seconds.
For the antenna design, GPT-4o-mini (16 seconds on average) and DeepSeek Chat v3.1 (13 seconds on average) incurred longer runtimes with broader distributions and higher upper outliers in contrast with o1 (12 seconds on average), Codestral 2508 (10 seconds on average) and GPT-4o (9 seconds on average).
The FSM example had stable runtimes. GPT-4o (2 seconds on average) and Codestral 2508 (3 seconds on average) models had relatively low runtimes, followed by GPT-4o-mini (4 seconds on average), while DeepSeek Chat v3.1 (9 seconds on average) and o1 (21 seconds on average) showed longer runtimes.

Figure \ref{Figure_5}b reports the LLM token count across the five selected LLM models.
In the OTA case, o1 required the highest number of tokens (1220 tokens on average), followed by Codestral 2508 (375 tokens on average), while GPT-4o (350 tokens on average), GPT-4o-mini (360 tokens on average) and DeepSeek Chat v3.1 (130 tokens on average) produced comparatively compact responses, with little variation across runs. 
For the antenna example, the o1 model required more tokens (1280 tokens on average) than Codestral 2508 (1100 tokens on average), GPT-4o-mini (930 tokens on average), GPT-4o (660 tokens on average) and DeepSeek Chat v3.1 (573 tokens on average).
For the FSM example, o1 exhibited the lowest token usage (107 tokens on average) compared to GPT-4o (140 tokens on average), GPT-4o-mini (150 tokens on average), DeepSeek Chat v3.1 (156 tokens on average) and Codestral 2508 (199 tokens on average).

Figure \ref{Figure_5}c shows the success rate across the five selected models. For each circuit, we considered the scenarios where the user prompt is equipped with design constraints and the case of a generic user prompt without design constraints.
In the OTA case, constrained runs showed consistently higher success rate, with GPT-4o-mini, GPT-4o, o1 and DeepSeek Chat v3.1 satisfying the constraints across all the five attempts. Under unconstrained conditions, o1 was the only model to fully satisfy the target, followed by GPT-4o. All models, except for GPT-4o-mini, showed successful netlist generation at least in one attempt.
In the antenna case, GPT-4o-mini showed strong success rate across all runs, followed by GPT-4o and o1 in terms of valid output generation. Under unconstrained conditions, GPT-4o-mini again showed the strongest performance, although this is mitigated with respect to the constrained case. Overall, the trend reflects the challenge of producing a working antenna netlist in ADS format, which is more tool-specific than a conventional Spice netlist or a Verilog script.
In the FSM case, constrained runs showed higher overall success, with GPT-4o-mini and GPT-4o satisfying the target in all five attempts. Under unconstrained conditions, none of the models satisfied the target, although GPT-4o, o1 and DeepSeek Chat v3.1 still produced valid Verilog scripts in all five attempts.
Overall, as reported in \ref{Figure_5}c, there are cases where a compatible Netlist or Verilog script were not properly generated. We thus investigated the reasons by parsing the reports and categorised the relevant errors as reported in Supplementary Table \ref{Table_S4}. In addition to the prompt-sensitivity study here described, Supplementary Tables \ref{Table_S12}, \ref{Table_S13}, \ref{Table_S14} detail ablation studies across the three designs to show the tool's behaviour when system prompts, technology binding logic, and EDA specifications are included or not. Overall, the results confirm that all the components are necessary to successfully simulate/implement the LLM-generated designs.

\section*{Conclusions}

We have reported LaMDA, a flexible framework that enhances EDA by integrating LLMs with commercial EDA tools for versatile circuit design across analogue, RF, and digital domains. Traditional EDA processes are crucial for complex circuit design, but often require multiple manual design iterations to meet the specifications, involving detailed log file analysis and PPA reports, which can be time-consuming and prone to errors. To address this, we proposed LaMDA, a framework that streamlines this design process by facilitating a collaborative human-LLM-EDA approach.
LaMDA enables two operational modes: custom design optimisation via LLM-assisted recommendations, and dataset benchmarking for LLM performance analysis. The framework generates EDA-compliant analogue and RF netlists or Verilog HDL scripts, compatible with tools such as Cadence Virtuoso Spectre, Keysight ADS, and AMD Vivado.
Evaluated through case studies on OTAs, microstrip patch antennas, and digital circuits, LaMDA acts as an interactive, EDA-aware assistant. It provides informed guidance to users with varying levels of expertise and helps addressing diverse design challenges.

\section*{Methods}

\subsection*{Analogue design flow setup}
We interfaced Cadence Spectre with the LaMDA framework for analogue circuit sizing, simulation and optimisation. All analogue simulations were run using Cadence Spectre (MMSIM 15.10.627; Spectre 15.1.0.627.isr12) with the TSMC 180 nm BCD models at typical-typical (TT) corner, supply voltage (VDD) = 5 V, and temperature (T) = 27 $^\circ$C. We used the OpenAI o1 for OTA analysis and GPT-4o-mini for inverter to generate Spectre-compatible simulation decks and to propose iterative design updates. The LLM models were configured to use the default values for $max\_tokens$, $temperature$, $top\_p$. The LLM is accessed via a Python API-based workflow. For the analogue examples, the LLM produces a raw Spectre-compatible simulation deck that combines a parametrised transistor-level netlist, stimuli, load conditions, and the required analysis directives. Importantly, PDK-dependent information (model inclusion, device mappings, corner sections for the TSMC 180 nm BCD kit) is applied during this local binding step after the raw LLM netlist is generated, producing a PDK-bound, Spectre-compatible netlist while keeping confidential PDK details hidden from the LLM.

To improve prompt efficiency and reproducibility, repetitive tool operations are not described inside the prompt. Instead, run-directory initialisation, Spectre invocation, and results parsing are implemented as reusable automation modules and executed programmatically, independently of the specific circuit. The system prompt is therefore restricted to design-dependent content needed for valid netlist generation, while execution logic remains in the framework. After each run, the framework parses Spectre PSF outputs and log files to extract key analogue metrics and to flag errors or critical warnings. 

The detailed system prompt used for the OTA design is reported in Supplementary Tables \ref{Table_S5}, \ref{Table_S6}, \ref{Table_S7}. In terms of EDA tool support, the prompts explicit support to Cadence Spectre. However, the user can customise the system prompt for different netlist-level simulators. PDK information is not embedded at the prompt level to preserve Intellectual Property (IP) confidentiality. The inclusion of the PDK-related specifications is performed afterwards. Specifically, a Python module post-processes the LLM-generated netlist, including PDK paths to model files and NMOS/PMOS device specifications. Eventually, it is worth underlining that the system prompt complies with any LLM provider.

For the 5T OTA case study, the flow enforces practical operating conditions by constraining the device operating points to avoid subthreshold operation, ensuring that the proposed solutions remain physically meaningful for the target technology. The framework successfully achieved the requested analogue targets. Notably, the optimisation outcome is prompt-dependent, as the specified objectives and constraints directly steer the parameter. The Python file updates the $V_{BIAS}$ across sweeps to met the target specifications. The $V_{BIAS}$ values were chosen by a script that randomly samples within a defined range, and we showed 15 sweeps to illustrate how the performance varies across different bias points. Additional specifications can be included if required. For example, tighter constraints on unity-gain bandwidth, power consumption, output swing, or noise can be added to the prompt and incorporated into the optimisation loop to further tailor the circuit behaviour. 

\subsection*{RF design flow setup}
We used OpenAI o1 model through a chat-based option, to generate Keysight ADS compliant netlists. The LLM model was configured to use the default values for $max\_tokens$, $temperature=1.0$, $top\_p=1.0$. We interfaced Keysight ADS EDA tool with LaMDA for RF Antenna design and simulation, specifically the Keysight ADS 2025 Update 0.1 version. A Python-based application programming interface (API) is available since the 2025 release\cite{Ozalas_2025} that allows for headless automation using external Integrated Development Environment (IDE), like Visual Studio Code. Custom Python functions can thus be used in scripts, running in the background, and eliminating the need for user interface (UI). However, information about this API is not yet widely accessible. Therefore, this information must be provided to the LLM.

The user prompt contained the user’s design description and objectives, expressed as target performance metrics. The system prompt was initially constructed to include exhaustive information on all available functions and operations within Keysight ADS. Although this approach ensured broad coverage, it resulted in a large prompt size and a very large token count with substantial information redundancy. To address this, frequently repeated functions such as workspace and library creation, cell and schematic generation, and simulation setup and execution, were fully automated to a global, reusable format. These operations are invoked programmatically and are independent of the specific design under consideration. Consequently, detailed descriptions of these functions were removed from the system prompt, as they no longer needed to be explicitly provided to the LLM. Differently, design-dependent information was retained. Specifically, a structured list of available Keysight ADS libraries and their corresponding components was extracted from Keysight ADS and included in the prompt to guide valid component selection. In addition, representative examples of Keysight ADS netlists were incorporated to provide syntactic and semantic reference patterns for netlist generation. Finally, the Keysight ADS Python API reference material supplied to the LLM was substantially reduced to include only the subset of functions required for netlist generation and iterative design modification. This prompt restructuring resulted in a significant reduction in token usage. The original prompt required approximately 128000 tokens, whereas the optimised prompt required approximately 26000 tokens, corresponding to a reduction of approximately 80\%. This reduction improved computational efficiency while maintaining the necessary information for accurate Keysight ADS-compatible netlist generation. Supplementary Table \ref{Table_S8} details the organisation of the full prompt, including system prompt details, ADS-related libraries, components, and simulator specifications, and example netlist. In addition, the feedback-loop user prompt format is reported. Furthermore, Supplementary Table \ref{Table_S9} specifies the structure and features of the ADS-related information.

LaMDA framework includes a pre-processing step to extract the netlist block from the LLM response and ensures tool-ready formatting by correcting basic syntax errors. A netlist represents a circuit in text form, but designers usually rely on schematics rather than netlists to evaluate designs\cite{matsuo2025schemato}. Access to a circuit diagram is therefore essential for verifying correct connectivity and for providing meaningful feedback to the LLM. Currently, Keysight ADS cannot automatically convert a netlist into a schematic. To overcome this, we proposed a graph-based netlist representation that acts as an intermediate circuit diagram and provides a foundation for error handling in complex designs using graph neural networks.

The design process followed an iterative human–LLM interaction. Therefore, an iteration is defined as one complete interaction cycle comprising (a) a user prompt to the LLM, (b) the LLM’s generated response, and (c) user evaluation and feedback. Using this process, 10 successive iterations were performed. After netlist simulation, the results are automatically plotted, allowing the user to assess whether the target performance is met.

Lastly, for the RF antenna design example, two additional results were subsequently produced by Keysight ADS: (a) the antenna layout and its physical dimensions and (b) the radiation pattern, obtained from an EM simulation in Keysight ADS, normalised to its maximum value and expressed in dB as a function of the polar angle $\theta$, to evaluate the antenna performance. From the EM simulation, the maximum directivity, maximum gain and efficiency for the frequency of interest ($f=2.4$ GHz), were also extracted.

\subsection*{FPGA design flow setup}
We used OpenAI GPT-4o to generate Verilog HDL scripts. The LLM model was configured as follows: $max\_tokens=3000$, $temperature=1.5$, $top\_p=0.75$, as in the ResBench work\cite{Guo_2025}. We interfaced the AMD Vivado 2024.1 EDA tool with LaMDA for FPGA design, simulation, synthesis, and implementation. The AMD Zynq 7020 FPGA ($xc7z020clg400-1$) was used to implement the designs.

Supplementary Table \ref{Table_S10} specifies the prompts managed by the LLM to generate the ResBench artifacts. The target LLM is required to be expert in FPGA design, Verilog and Python coding, with no specific reference to the EDA tool responsible for simulating and implementing the design. Eventually, the prompt format is compliant with any LLM provider.

The ResBench dataset was organised within a JavaScript Object Notation (JSON) file. From the original dataset, each problem included the following items (example in Supplementary Fig. \ref{Figure_FS5}a): top module name; problem definition; module header, including input and output ports; Verilog HDL testbench for simulation.
Before the analysis, we processed the JSON file by adding the following specifications (example in Supplementary Fig. \ref{Figure_FS5}b): problem ID (following the scheme in Supplementary Table \ref{Table_S1}); LUT count objective; maximum delay objective (to be used with fully-combinatorial logic); clock frequency objective (to be used in the case of pipelined designs); clock constraint attribute to be included into the FPGA constraint file read by AMD Vivado.

To determine the LUT count objective per problem, we considered the minimum LUT count achieved by the ResBench work\cite{Guo_2025} across all the validated LLMs. Regarding the maximum delay, we randomised the selection between 1 and 10 ns. Similarly, we randomised the clock frequency between 100 MHz and 1000 MHz, with steps of 50 MHz.

For each run, we collected key output results into a JSON file that includes the following items (example in Supplementary Fig. \ref{Figure_FS6}a): problem ID; top module name; clock constraint; Verilog HDL script generated by the LLM; LLM token count; LLM generation time; AMD Vivado execution time from design instantiation to implementation; logic simulation outcome; synthesis outcome; implementation outcome; resource count, detailing the number of LUTs, registers, on-chip block random access memories (BRAMs), digital signal processing (DSP) elements, input/output ports (I/O); delays, including data path delay, logic delay, route delay; power consumption, including total power, and dynamic and static power fractions; LUT count objective outcome; maximum delay/clock frequency outcome.
In the case of unsuccessful designs, the JSON file also includes key error logs (example in Supplementary Fig. \ref{Figure_FS6}b).

\subsection*{Large language models evaluation setup}

We evaluated the three case studies across five LLM models: OpenAI GPT-4o, GPT-4o-mini and o1; Codestral 2508; and DeepSeek Chat v3.1. Each case was repeated five times to support statistical comparison. We compared LLM response time, output token count and success rate.
To keep antenna reporting consistent with the OTA and FSM cases, we considered the earliest iteration that meets the design target (selected iterations reported in Supplementary Table \ref{Table_S11}); if the target was not met overall, we considered the iteration with the best \%Var (as reported below); in case of failed iterations, we selected the results from the last one for LLM response time and output token count.
\begin{equation}
    \%Var = \frac{|S_{11, Obtained}|-|S_{11, Target}|}{|S_{11, Target}|}\times100
\end{equation}

\subsection*{Code availability}
The supported LaMDA software is open-sourced and available at the following link: \url{https://github.com/aprilaihub/LaMDA}.

\section*{Acknowledgments}
We thank EDINA at the University of Edinburgh for supporting the access to the OpenAI LLMs through the ELM gateway.

\section*{Funding}

This work was supported by the Engineering and Physical Sciences Research Council (EPSRC) AI Hub for Productive Research and Innovation in eLectronics (APRIL) under Grant No. EP/Y029763/1, and by the Royal Academy of Engineering (RAEng) Chair in Emerging Technologies under Grant No. CiET1819/2/93.

\section*{Author contributions}

C.S., A.D., P.V., P.K. realised the framework: P.V. was responsible for the analogue section; P.K. was responsible for the RF section; C.S. and A.D. were responsible for the FPGA section. P.V. ran the experiments and data collection for the OTA and inverter case studies. P.K. ran the experiments and data collection for the microstrip patch antenna design. C.S. and A.D. ran the experiments and data collection for the FPGA dataset. A.K. implemented the graph-based visualisation of the RF netlist and assisted in the graph generation across the design iterations. C.S., P.K., P.V. wrote the draft manuscript. All authors reviewed the manuscript and agreed on the final version. M.O.B., C.S.B., T.P. supervised the work and provided critical comments throughout the different stages. 

\section*{Competing interests}

The authors declare no conflict of interest.

\section*{Data Availability}
The dataset generated during the current study is available at \url{https://doi.org/10.7488/ds/8199}.

\nolinenumbers

\bibliographystyle{naturemag}
\bibliography{Sections/10_References}

\clearpage

\setcounter{page}{1}

\title{A flexible language model-assisted \\electronic design automation framework\\\textbf{Supplementary Information}}

\author{
    Cristian Sestito\textsuperscript{1,*},
    Panagiota Kontou\textsuperscript{1},
    Pratibha Verma\textsuperscript{1,2}, \\
    Atish Dixit\textsuperscript{1},
    Alexandros D. Keros\textsuperscript{1},\\
    Michael O'Boyle\textsuperscript{3},
    Christos-Savvas Bouganis\textsuperscript{4},
    Themis Prodromakis\textsuperscript{1}
}

\date{
    \small
    \textsuperscript{1}Centre for Electronics Frontiers, Institute for Integrated Micro and Nano Systems, \\ School of Engineering, The University of Edinburgh, UK \\
    \textsuperscript{2}Department of Electrical Engineering, Indian Institute of Technology Indore, India \\
    \textsuperscript{3}School of Informatics, The University of Edinburgh, UK \\
    \textsuperscript{4}Department of Electrical and Electronic Engineering, Imperial College London, UK \\[0.2in]
    \textsuperscript{*}Correspondence: csestito@ed.ac.uk \\
}

\maketitle

\renewcommand{\thefigure}{S\arabic{figure}}
\setcounter{figure}{0} 

\renewcommand{\thetable}{S\arabic{table}}
\setcounter{table}{0} 

\begin{figure}
\includegraphics[width=\textwidth]{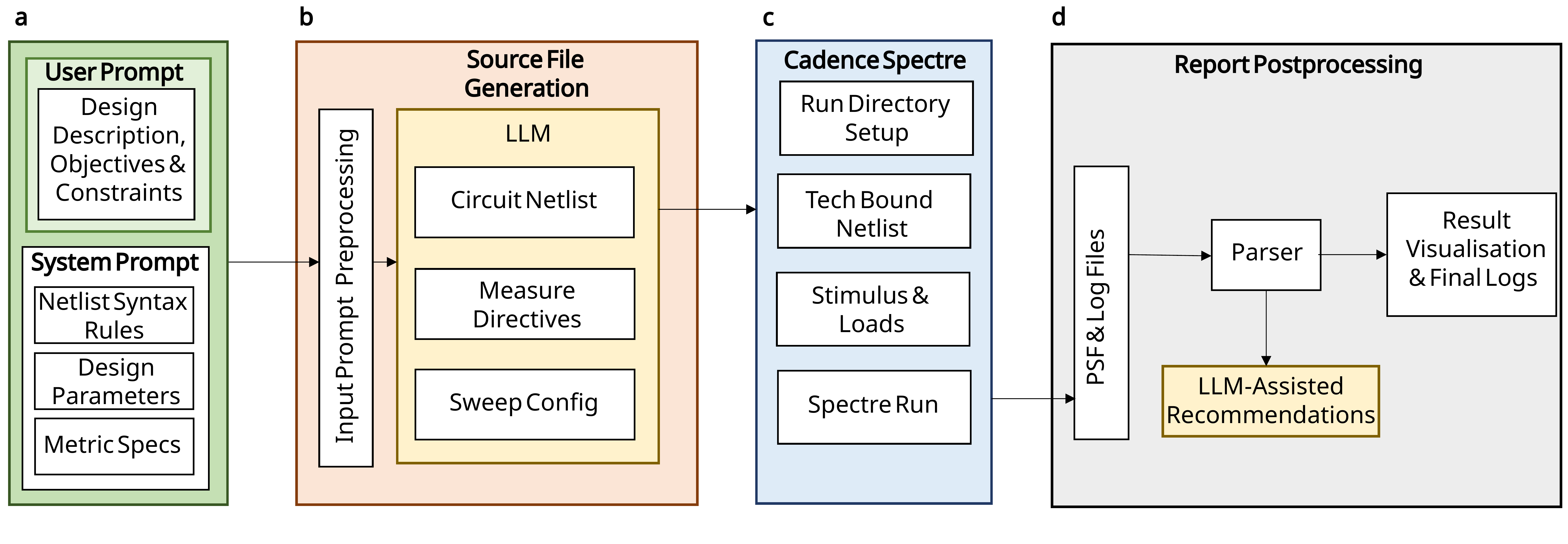}
\centering
\caption{
\textbf{A block diagram of the LaMDA framework running the analogue design flow.} 
\textbf{a,} The input prompt contains the user’s design description and objectives, expressed as target performance metrics (for example, DC operating conditions, AC stability and bandwidth metrics, and transient response requirements) together with run conditions and constraints (such as supply, load, and allowable parameter ranges). A system prompt provides technology context, including technology information and required simulation stimuli.
\textbf{b,} The Source File Generation module includes an LLM that generates the circuit netlist and analysis setup files (DC/AC/transient), and produces constraint and parameter definitions suitable for automated exploration. Input-prompt preprocessing structures the specifications and constraints into a Spectre tool-ready format.
\textbf{c,} The Cadence Virtuoso Spectre tool reads the generated source files, creates a workspace, runs netlist, and perform circuit simulations, producing waveform databases (.psf) and log files (.log). 
\textbf{d,} The Report Postprocessing module parses the simulation outputs to extract the key metrics, generates structured logs and visualisations, and checks whether the objectives are met.}
\label{Figure_AS1}
\end{figure}

\begin{figure}[t]
\includegraphics[width=\textwidth]{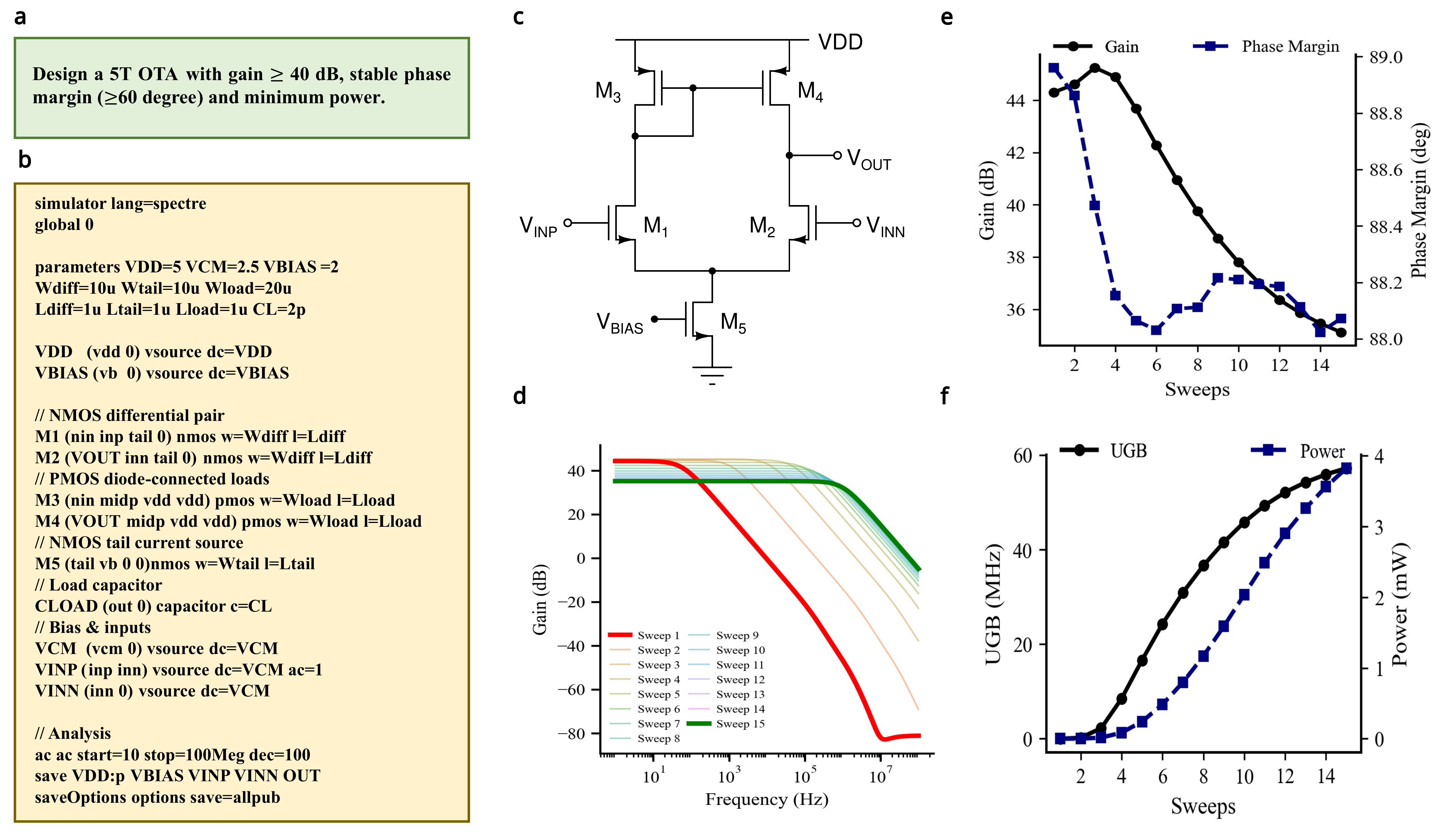}
\centering
\caption{
\textbf{Analogue design flow example: OTA with generic design specifications.} 
\textbf{a,} Input prompt. We require an OTA, having a gain of at least 40\,dB and with a stable phase margin ($\geq$ $60^\circ$).
\textbf{b,} Generated netlist. The netlist includes information about the device parameters ($W_{\mathrm{diff}}$, $L_{\mathrm{diff}}$, etc.) and includes necessary analysis commands and voltage sources, demonstrating the LLM's capability to generate functional SPICE code. The generated netlist is compatible with Cadence Virtuoso Spectre. 
\textbf{c,} The 5-transistor OTA circuit with an NMOS differential pair (M1, M2), PMOS diode-connected loads (M3, M4), and an NMOS tail current source (M5).
\textbf{d,} OTA gain from simulations. The graph shows the reduced DC gain and increased UGB with increasing bias, illustrating the speed--gain trade-off. 
\textbf{e,} DC gain and phase margin across 15 sweeps. Gain decreases with increasing tail current, while PM remains near $90^\circ$, confirming stability. 
\textbf{f,} UGB across 15 sweeps. UGB improves with iteration, but at the cost of steadily increasing power,
with bandwidth gains gradually saturating at later sweeps.}
\label{Figure_AS2}
\end{figure}

\begin{figure}
\includegraphics[width=\textwidth]{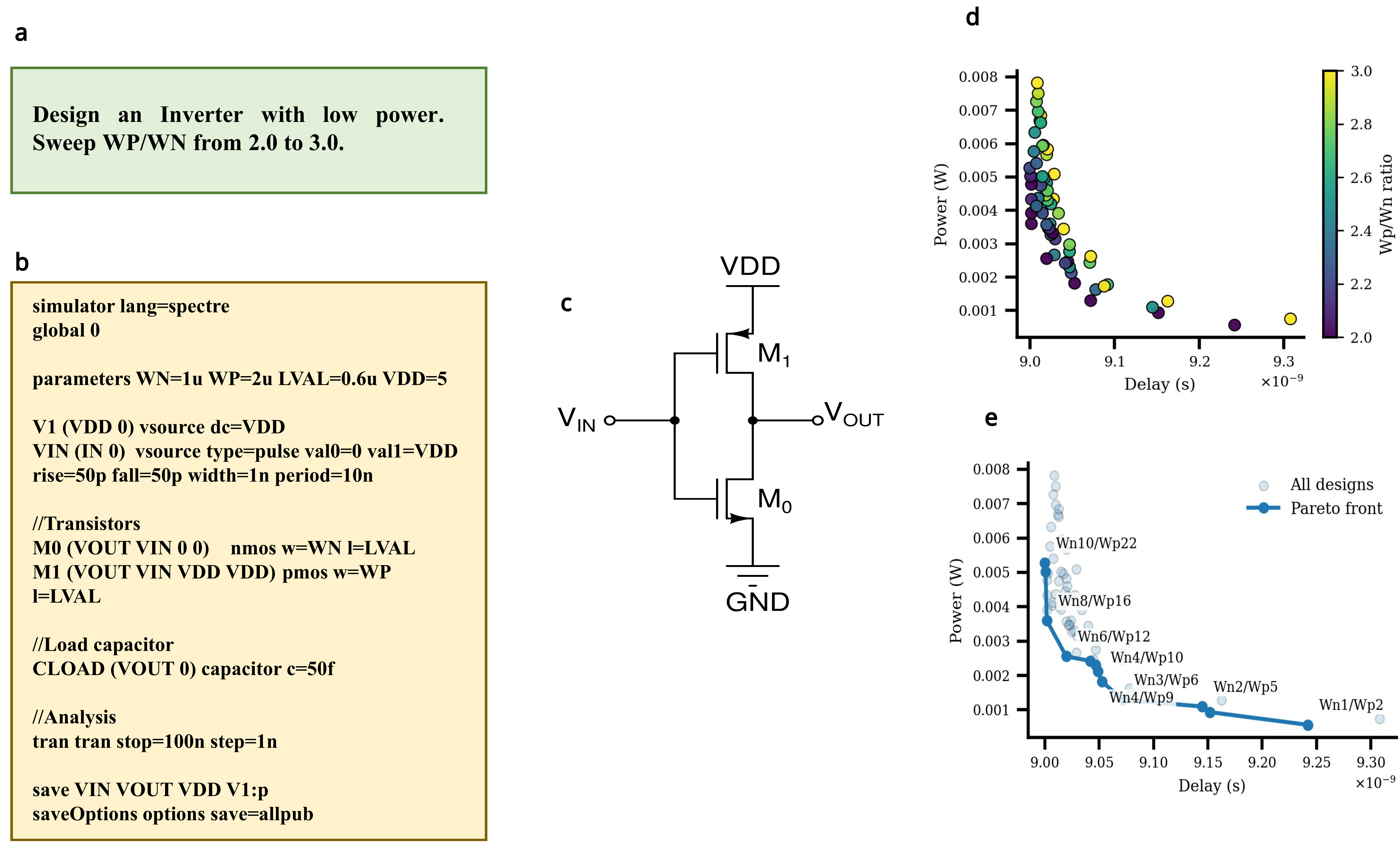}
\centering
\caption{
\textbf{Analogue design flow example: Inverter with generic design specifications.} 
\textbf{a,} Input prompt. The user requests a CMOS inverter optimised for low power and with a target WP/WN ratio.
\textbf{b,} Generated netlist. The LLM produces a functional Spectre-compatible SPICE netlist, including device sizing parameters, stimulus sources, and required analysis commands.
\textbf{c,} Inverter schematic. The resulting CMOS inverter consists of a PMOS pull-up and NMOS pull-down pair connected at the output node. 
\textbf{d,} Delay–power trade-off from simulations. A parameter sweep across transistor sizing identifies the relationship between power consumption and switching delay. 
\textbf{e,} Pareto front for power vs worst-case delay ($tp_{HL}$). The non-dominated designs highlight the best achievable trade-offs, revealing sizing points that minimise power for a given worst-case propagation delay.}
\label{Figure_AS3}
\end{figure}

\begin{figure}
\includegraphics[width=\textwidth]{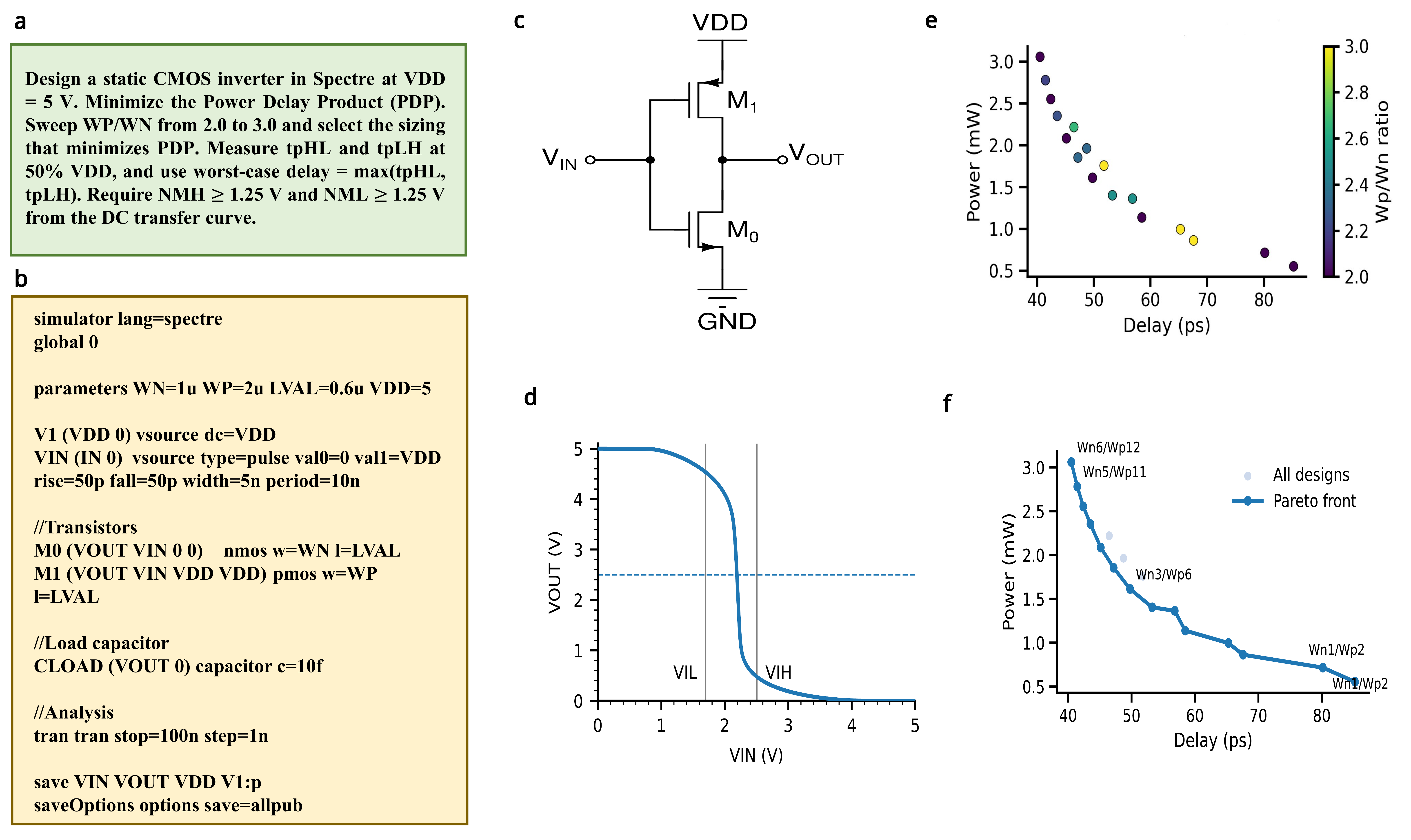}
\centering
\caption{
\textbf{Analogue design flow example: Inverter with detailed design specifications.} 
\textbf{a,} Input prompt. The user requests a static
CMOS inverter in Spectre at VDD = 5\,V and specifies minimisation of power–delay product (PDP) with noise-
margin constraints. 
\textbf{b,} Generated netlist. The LLM produces a functional Spectre-compatible SPICE netlist.
\textbf{c,} Inverter schematic. The resulting CMOS inverter consists of a PMOS pull-up and NMOS pull-down pair connected at the output node. 
\textbf{d,} Noise margins from DC transfer curve. The extracted NMH and NML values are used to
verify functional robustness, enforcing NMH $\geq$ 1.25\,V and NML $\geq$ 1.25\,V for the selected designs.The LaMDA
framework performs 15 iterations to reach the design objectives. 
\textbf{e,} Delay–power trade-off from simulations. A
constrained parameter sweep across WP/WN identifies the relationship between power consumption and worst-case
switching delay computed as max ($tp_{HL}$, $tp_{LH}$) at 50\% VDD. 
\textbf{f,} Pareto front for power vs worst-case delay. The non-
dominated designs highlight the best achievable trade-offs under the specified objective and constraints, identifying
sizing points that minimise PDP while maintaining valid delay and functionality.}
\label{Figure_AS4}
\end{figure}

\begin{figure}
\includegraphics[width=\textwidth]{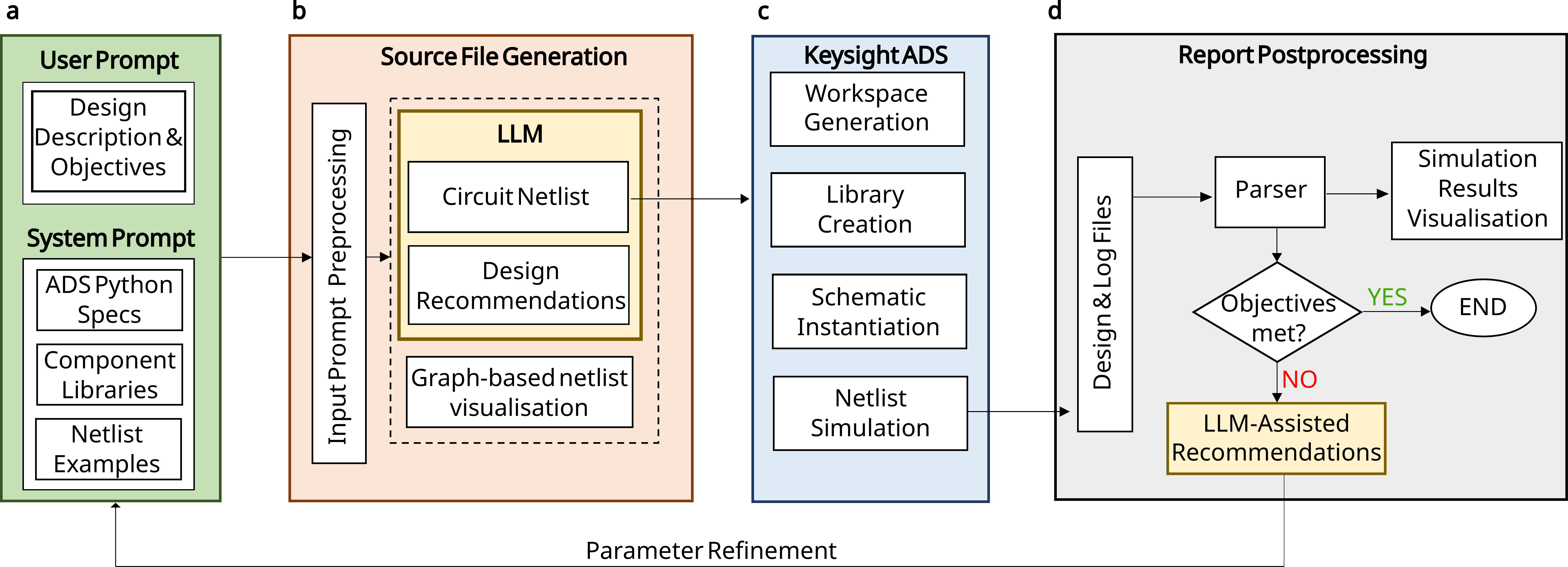}
\centering
\caption{
\textbf{A block diagram of the LaMDA framework running the radio-frequency design flow.} 
\textbf{a,} The input prompt contains the user’s design description and objectives, expressed as target performance metrics. The user prompts the LLM via a chat-based interface. A system prompt provides software-specific information such as available components and libraries, examples and Keysight ADS Python specification.
\textbf{b,} The Source File Generation module includes a LLM that generates the circuit netlist compatible with Keysight ADS format and provides design recommendations. Input-prompt preprocessing ensures correct structure and resolves potential errors, preparing the netlist for tool-ready use. Additionally, a graph-based visualisation of the netlist is produced, to validate correct connectivity between components.
\textbf{c,} The Keysight ADS tool creates a workspace, a library and a schematic to which the netlist is associated. The netlist is simulated producing results (.ds) (in this case reflection coefficient analysis) and log files (.log). 
\textbf{d,} The Report Postprocessing module parses the simulation results to generate visualisations and the user evaluates whether the design objectives are met. The framework supports two execution paths: termination of the process or continuation through additional optimisation cycles with LLM-assisted recommendations. In either case, the objectives specified in the input prompts are evaluated.}
\label{Figure_RS0}
\end{figure}

\begin{figure}
\includegraphics[width=\textwidth]{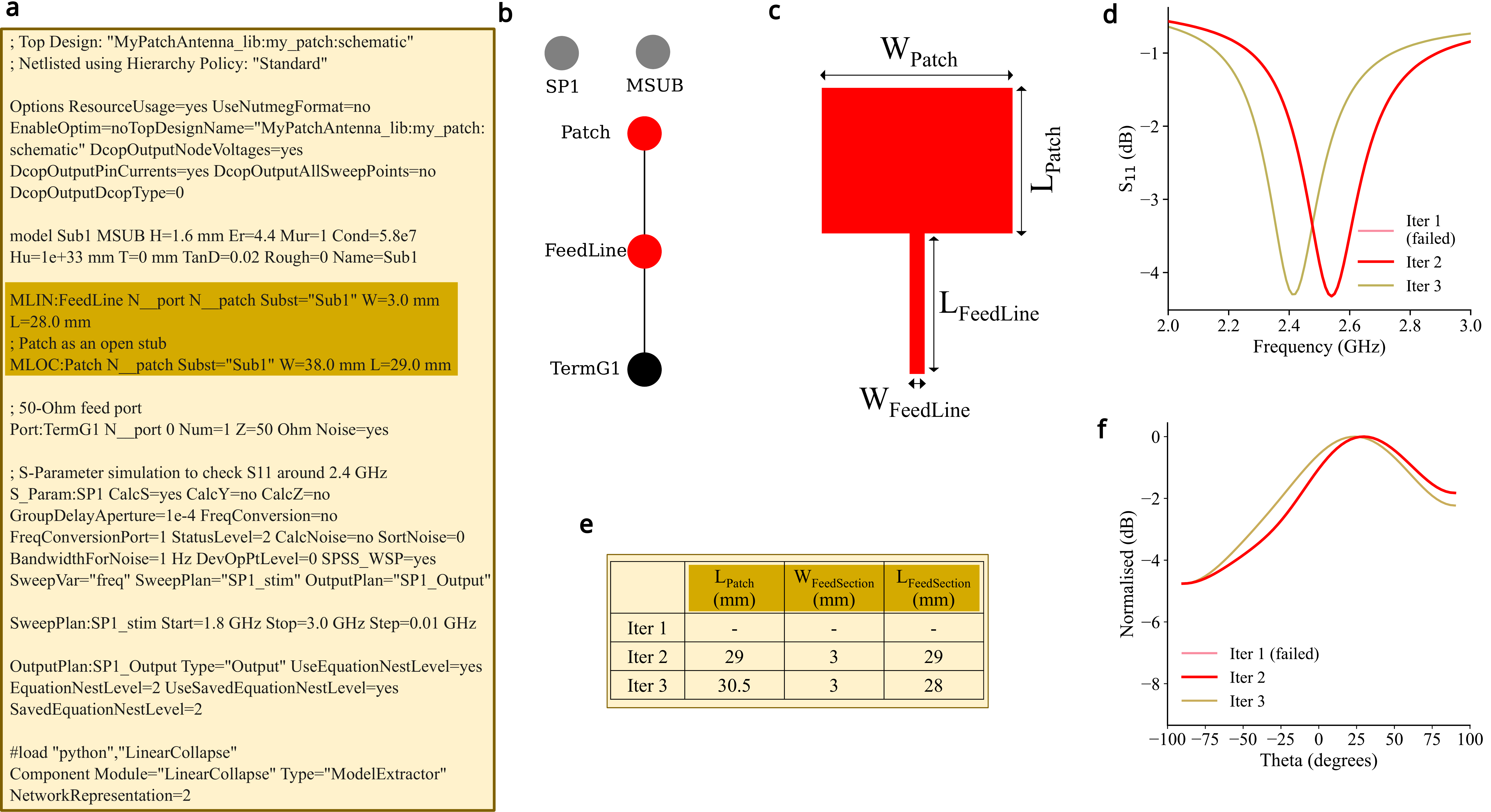}
\centering
\caption{
\textbf{RF design flow example: Iterations 1-3.} 
\textbf{a,} Netlist generated by LLM. The highlighted part denotes the component (feedline) that changes in width and length across iterations 1-3.
\textbf{b,} Graph representation of the netlist.
\textbf{c,} The equivalent antenna layout obtained from Keysight ADS. 
\textbf{d,} Table summarising the patch length and feedline length and width values for iterations 1-3.
\textbf{e,} Reflection coefficient from the 2D S-parameter simulation, showing improvement across iterations.
\textbf{f,} Radiation pattern associated with the antenna from a 3D EM simulation. 
}
\label{Figure_RS1}
\end{figure}

\begin{figure}
\includegraphics[width=\textwidth]{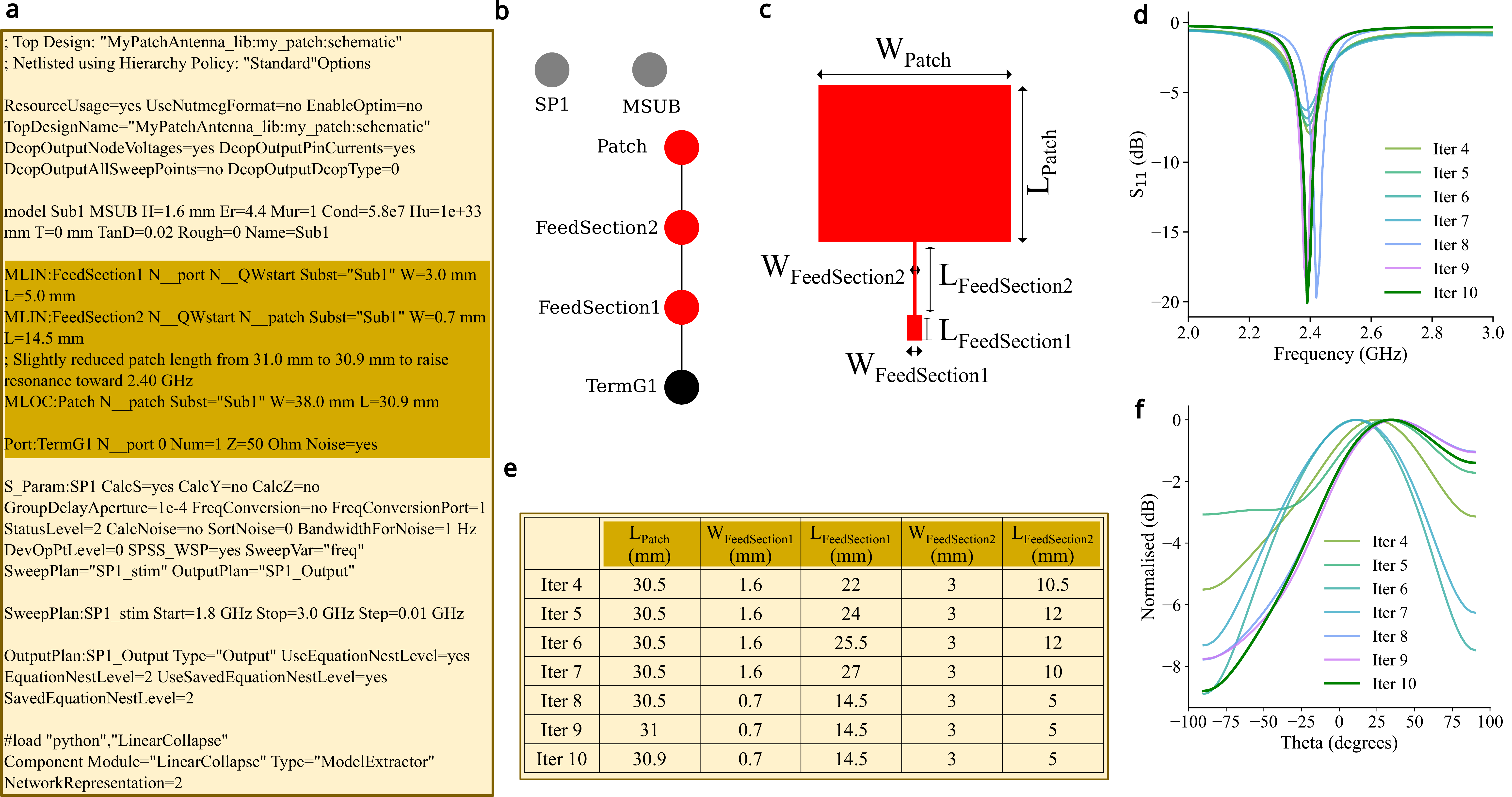}
\centering
\caption{
\textbf{RF design flow example: Iterations 4-10.} 
\textbf{a,} Netlist generated by LLM. The highlighted part denotes the component (feedline) that changes in width and length across iterations 4-10.
\textbf{b,} Graph representation of the netlist.
\textbf{c,} The equivalent antenna layout obtained from Keysight ADS. 
\textbf{d,} Table summarising the patch length and double feedline lengths and widths values for iterations 4-10.
\textbf{e,} Reflection coefficient from the 2D S-parameter simulation, showing improvement across iterations.
\textbf{f,} Radiation pattern associated with the antenna from a 3D EM simulation. 
}
\label{Figure_RS2}
\end{figure}

\begin{figure}
\includegraphics[width=\textwidth]{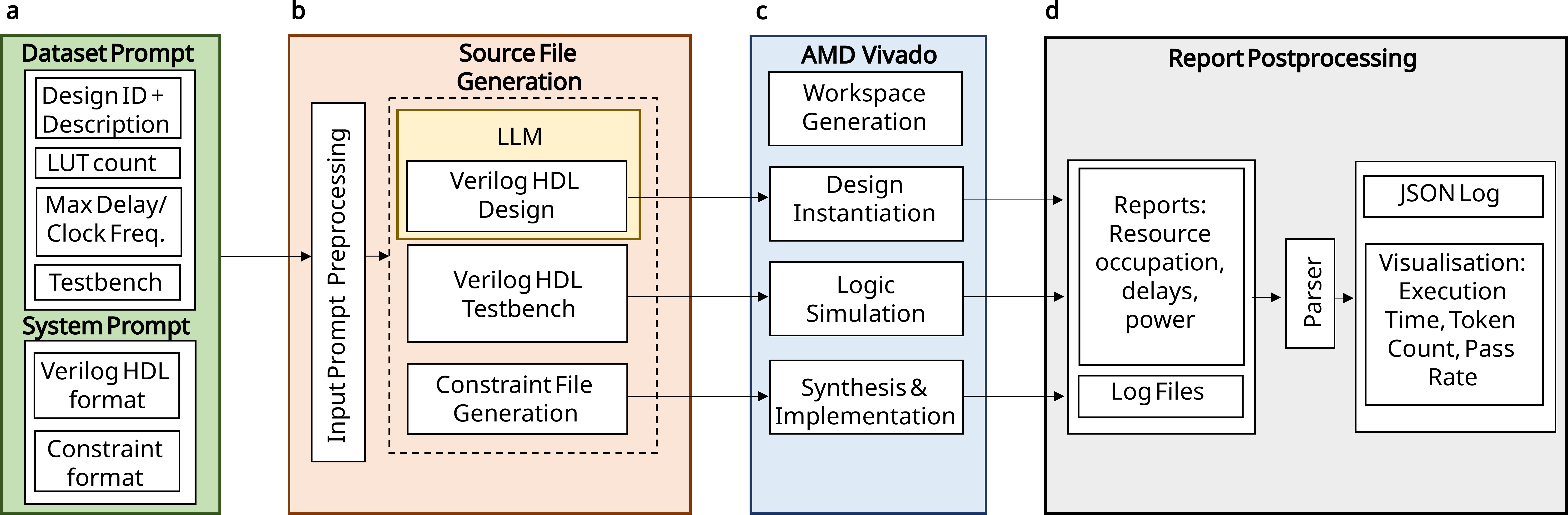}
\centering
\caption{
\textbf{A block diagram of the LaMDA framework running the FPGA design flow.} 
\textbf{a,} The user prompt comes from a dataset and includes design ID, description and objectives (i.e., LUT count and maximum delay/clock frequency). A testbench in Verilog HDL is provided as well. The system prompt provides syntax rules for Verilog HDL and constraint compliance with the EDA tool.
\textbf{b,} The Source File Generation module includes an LLM that generates the Verilog HDL design script for the AMD Vivado tool. Other source files, including the testbench and design constraints, are extracted from the prompt through the Input Prompt Preprocessing module.
\textbf{c,} The AMD Vivado tool reads the source files from the LLM, generates the workspace, instantiates the design, runs simulations, and executes the design synthesis and implementation. 
\textbf{d,} The PPA reports and log files are supplied to the Report Postprocessing module. Key log message and information, including resource occupation, delay, and power dissipation are extracted through a parser. LLM response time, full execution time, LLM response output token count, implementation pass rate are eventually visualised for further data analysis.}
\label{Figure_FS1}
\end{figure}

\begin{figure}
\includegraphics[width=\textwidth]{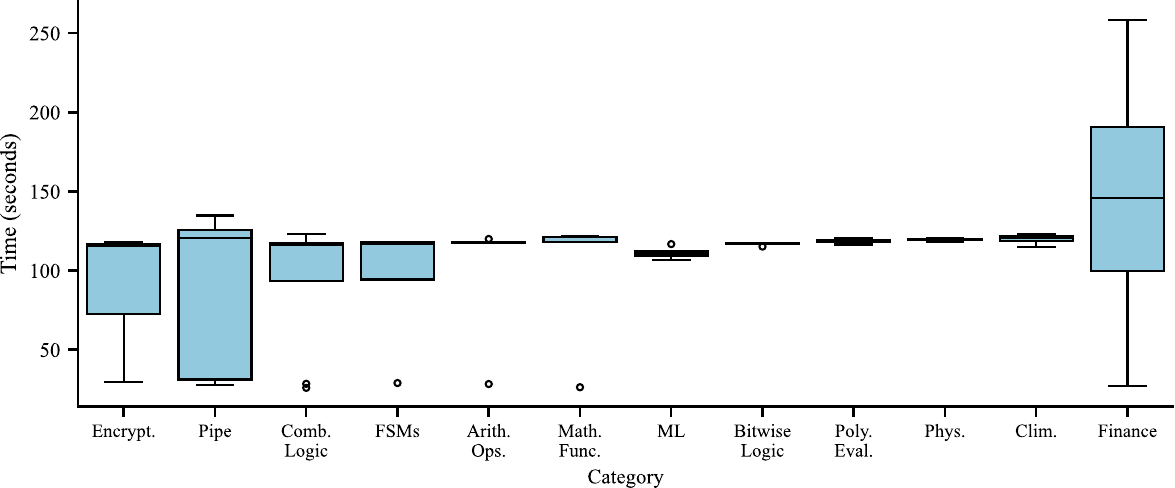}
\centering
\caption{
\textbf{FPGA design flow example: Total execution time analysis based on the ResBench dataset.} 
The total execution time includes the LLM response time, the AMD Vivado execution time (from design instantiation to implementation), and any overhead associated to software modules ran in-between. On average, the total execution time spans from 88 seconds (category: encryption) to 144 seconds (category: financial computing).
}
\label{Figure_FS2}
\end{figure}

\begin{figure}
\includegraphics[width=\textwidth]{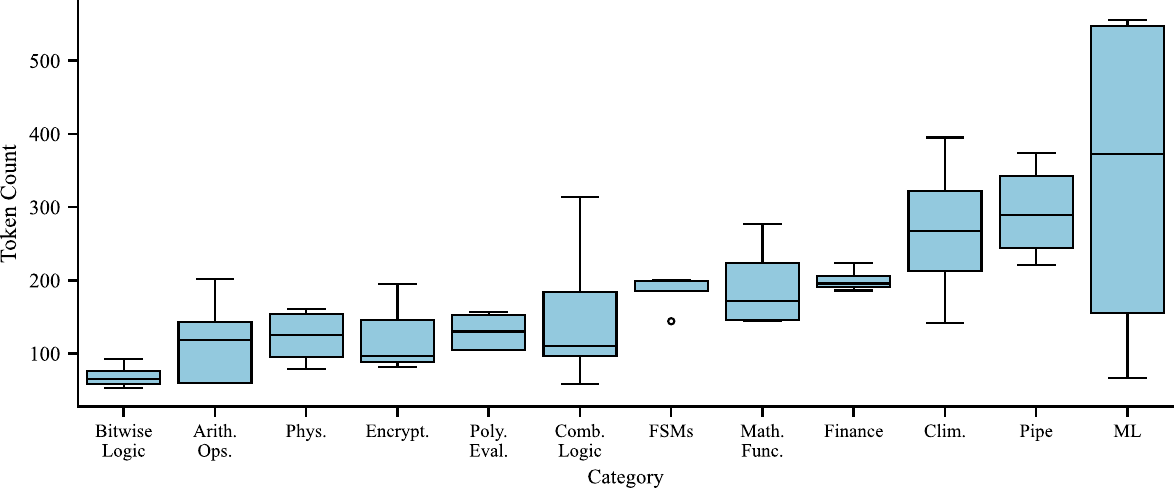}
\centering
\caption{
\textbf{FPGA design flow example: LLM token count analysis based on the ResBench dataset.} 
On average, the count spans from 69 tokens (category: bitwise and logical operations) to 340 tokens (category: machine learning).
}
\label{Figure_FS3}
\end{figure}

\begin{figure}
\includegraphics[width=\textwidth]{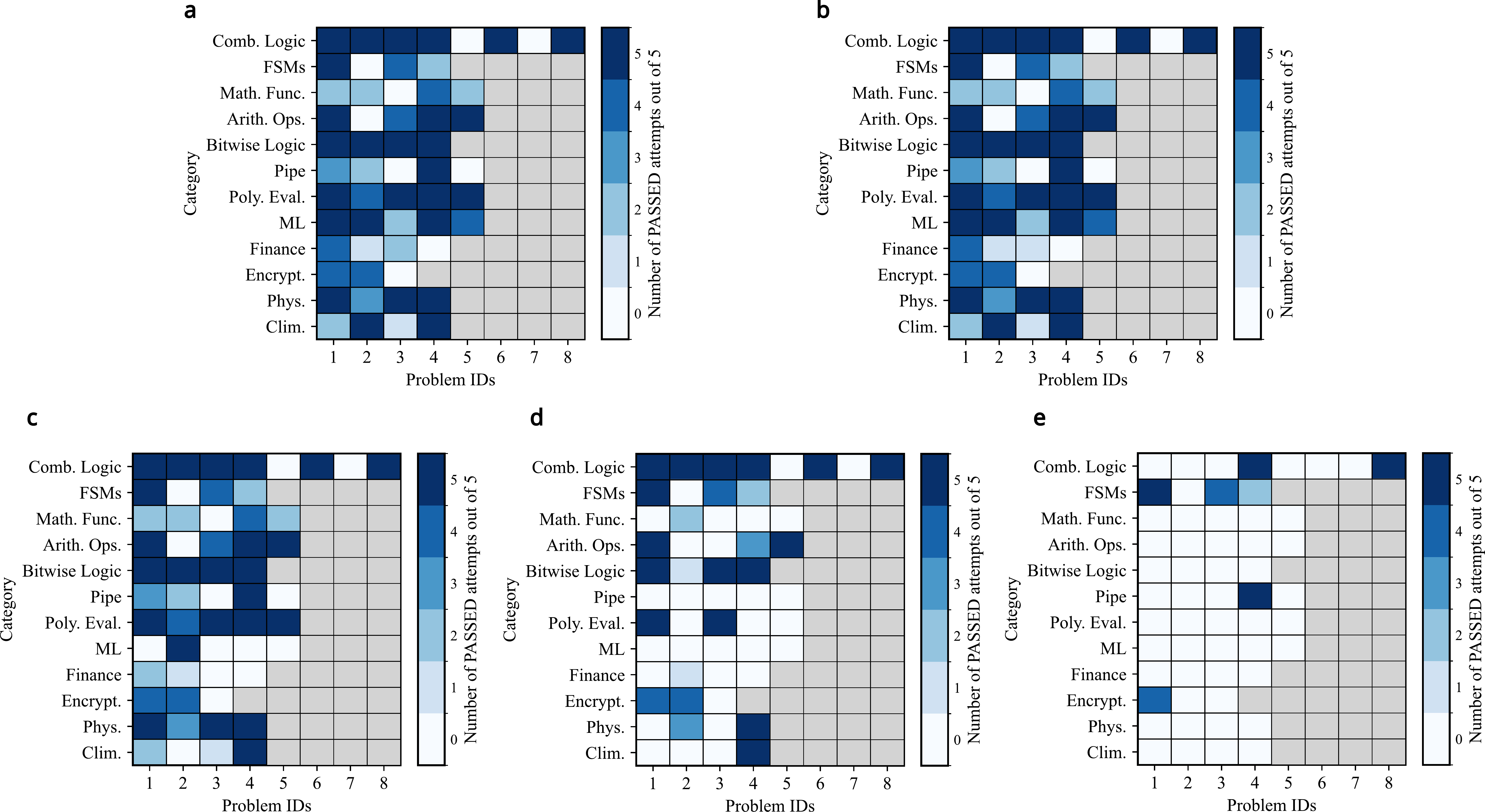}
\centering
\caption{
\textbf{FPGA design flow example: Pass rate analysis based on the ResBench dataset.}
\textbf{a,} Simulation pass rate.
\textbf{b,} Synthesis pass rate.
\textbf{c,} Implementation pass rate.
\textbf{d,} LUT objective pass rate.
\textbf{e,} Delay/Clock frequency objective pass rate.
}
\label{Figure_FS4}
\end{figure}

\begin{figure}
\includegraphics[width=\textwidth]{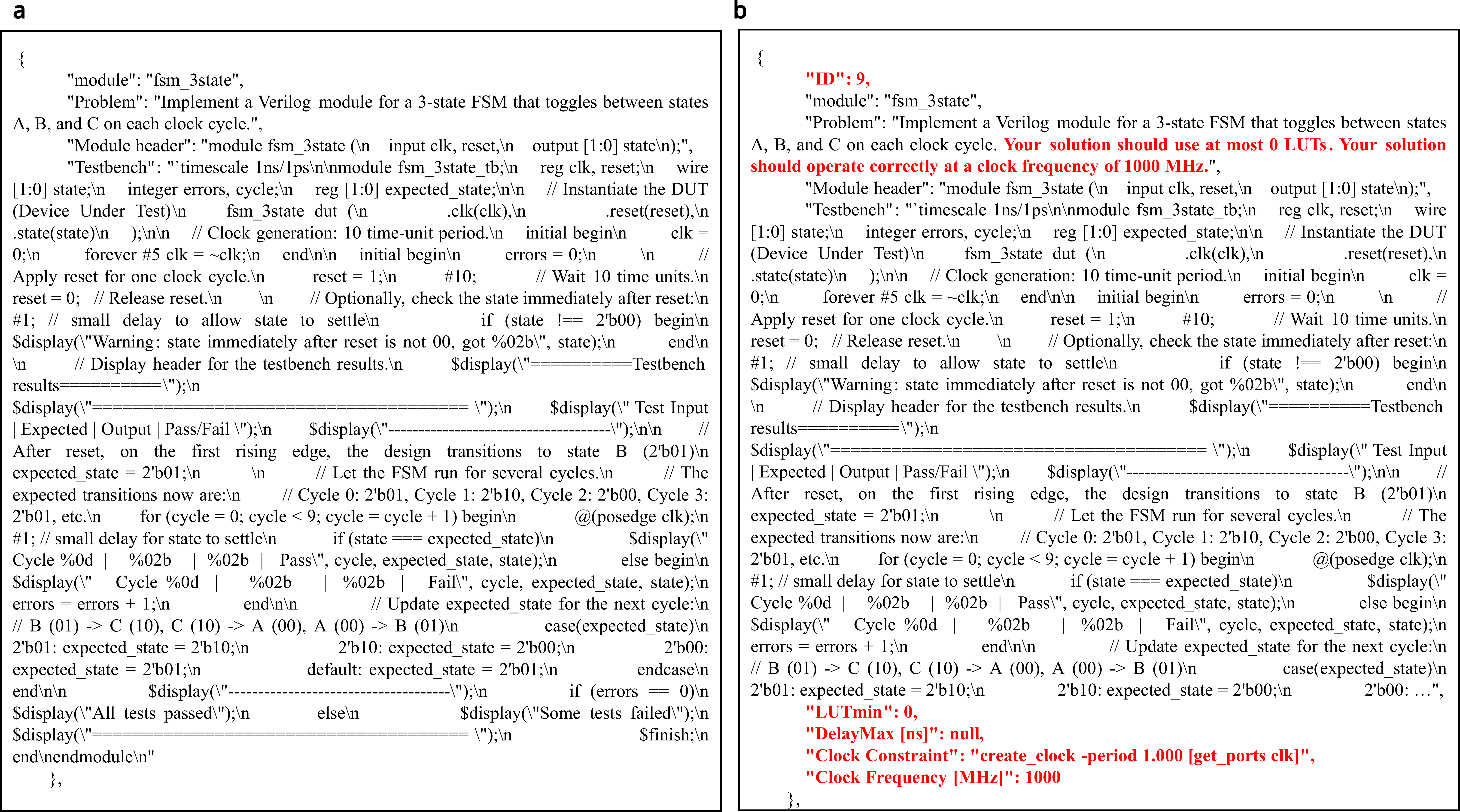}
\centering
\caption{
\textbf{FPGA design flow example: Example of input prompt from the ResBench dataset.}
\textbf{a,} Original prompt as from dataset.
\textbf{b,} Revised prompt including problem ID and design objectives (i.e., LUT count and clock frequency). Edits, automatically implemented by the LaMDA framework, are highlighted in red colour.
}
\label{Figure_FS5}
\end{figure}

\begin{figure}
\includegraphics[width=\textwidth]{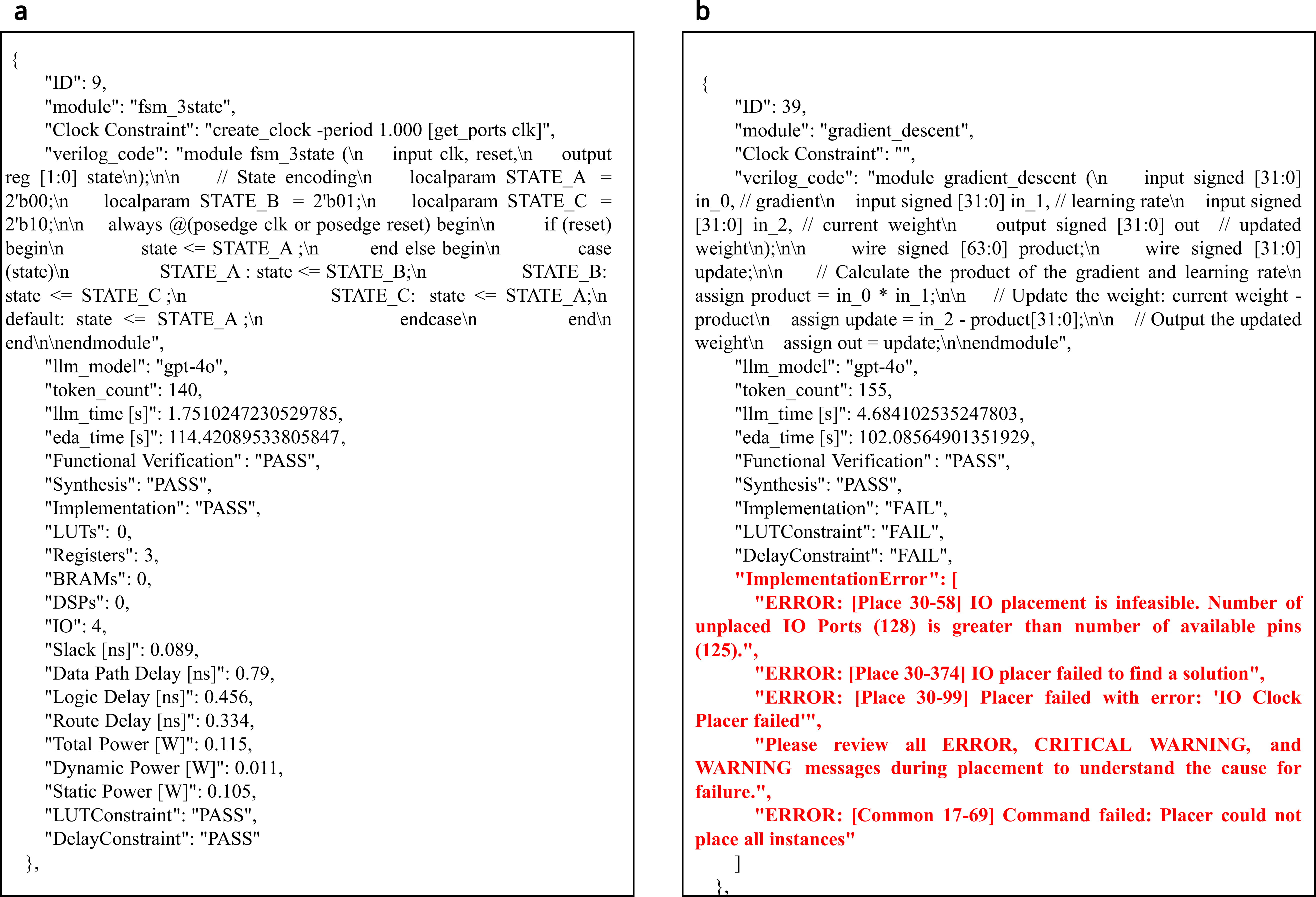}
\centering
\caption{
\textbf{FPGA design flow example: Examples of output JSON log.}
\textbf{a,} Module: 3-state FSM. Information including module name, LLM-generated Verilog design, LLM model, execution time, pass/fail outcomes for each stage, PPA information, pass/fail outcomes for design objectives are reported.
\textbf{b,} Module: gradient descent unit. As the implementation step fails, errors are also logged (highlighted in red colour).
}
\label{Figure_FS6}
\end{figure}

\begin{table}[t]
    \centering
    \caption{The ResBench Dataset: Categories and Problems}
    \label{Table_S1}
    \footnotesize
    \setlength{\tabcolsep}{4pt}
    \renewcommand{\arraystretch}{0.8}
    \begin{tabular}{c c c c}
    \hline
    Category & Abbreviation & Problem ID & Module \\
    \hline
    \multirow{8}{*}{Combinational Logic} & \multirow{8}{*}{Comb. Logic} & 1 & parity 8bit \\
     &  & 2 & mux4to1 \\
     &  & 3 & majority \\
     &  & 4 & bin to gray \\
     &  & 5 & eq comparator \\
     &  & 6 & decoder 2to4 \\
     &  & 7 & seven segment decoder \\
     &  & 8 & priority encoder \\
    \hline
    \multirow{4}{*}{Finite State Machines} & \multirow{4}{*}{FSMs} & 9 & fsm 3state \\
     &  & 10 & traffic light \\
     &  & 11 & elevator controller \\
     &  & 12 & vending machine \\
    \hline
    \multirow{5}{*}{Mathematical Functions} & \multirow{5}{*}{Math. Func.} & 13 & int sqrt \\
     &  & 14 & fibonacci \\
     &  & 15 & mod exp \\
     &  & 16 & power \\
     &  & 17 & log2 int \\
    \hline
    \multirow{5}{*}{Basic Arithmetic Operations} & \multirow{5}{*}{Arith. Ops.} & 18 & add 8bit \\
     &  & 19 & mult 4bit \\
     &  & 20 & abs diff \\
     &  & 21 & modulo op \\
     &  & 22 & subtract 8bit \\
    \hline
    \multirow{4}{*}{Bitwise and Logical Operations} & \multirow{4}{*}{Bitwise Logic} & 23 & bitwise ops \\
     &  & 24 & left shift \\
     &  & 25 & bitwise not \\
     &  & 26 & rotate left \\
    \hline
    \multirow{5}{*}{Pipelining} & \multirow{5}{*}{Pipe} & 27 & pipelined adder \\
     &  & 28 & pipelined multiplier \\
     &  & 29 & pipelined accumulator \\
     &  & 30 & pipelined max finder \\
     &  & 31 & pipelined fir \\
    \hline
    \multirow{5}{*}{Polynomial Evaluation} & \multirow{5}{*}{Poly. Eval.} & 32 & $x^2 + 2x + 1$ \\
     &  & 33 & $x^3 + 3x^2 + 3x + 1$ \\
     &  & 34 & $x^2 - x - 6$ \\
     &  & 35 & $(x+2)^2 + (x+2)^2 + (x+2)^2$ \\
     &  & 36 & $(a+b)^2 - (a-b)^2$ \\
    \hline
    \multirow{5}{*}{Machine Learning} & \multirow{5}{*}{ML} & 37 & matrix vector mult \\
     &  & 38 & relu \\
     &  & 39 & gradient descent \\
     &  & 40 & mse loss \\
     &  & 41 & conv2d \\
    \hline
    \multirow{4}{*}{Financial Computing} & \multirow{4}{*}{Finance} & 42 & compound interest \\
     &  & 43 & ddm \\
     &  & 44 & present value \\
     &  & 45 & currency converter \\
    \hline
    \multirow{3}{*}{Encryption} & \multirow{3}{*}{Encrypt.} & 46 & caesar cipher \\
     &  & 47 & modular add cipher \\
     &  & 48 & feistel cipher \\
    \hline
    \multirow{4}{*}{Physics} & \multirow{4}{*}{Phys.} & 49 & free fall distance \\
     &  & 50 & kinetic energy \\
     &  & 51 & potential energy \\
     &  & 52 & wavelength \\
    \hline
    \multirow{4}{*}{Climate} & \multirow{4}{*}{Clim.} & 53 & carbon footprint \\
     &  & 54 & heat index \\
     &  & 55 & air quality index \\
     &  & 56 & solar radiation average \\
    \hline
    \end{tabular}
    \end{table}

\begin{table}[t]
\centering
\caption{Details on LaMDA About Output Checks and Error Handling}
\label{Table_S2}
\small
\setlength{\tabcolsep}{5pt}
\renewcommand{\arraystretch}{1.2}
\begin{tabular}{p{0.24\linewidth} p{0.70\linewidth}}
\hline
Feature & Detail  \\
\hline
Output format checks & Before launching the specific design flows, LaMDA performs fixed preprocessing checks. In the analogue flow, LaMDA adds required headers for PDK compliance; in the FPGA flow, LaMDA extracts code from expected markers; in the RF flow, LaMDA extracts the netlist block from the model response. \\
\hline
Error tracking & In the different flows, LaMDA saves logs, status outputs, and intermediate artifacts (prompts, generated text, and EDA tool outputs), which supports diagnosis after runs. In the RF iterative mode, simulation feedback is also saved to guide the next generation step. \\
\hline
Safe tool execution & Runtime wrappers check environment variables/tool paths and execute backend EDA tools through controlled subprocess calls. \\
\hline
Netlist structure graph & In the RF flow, LaMDA includes a utility that converts a netlist into a connectivity graph, enabling structural inspection of components, nets, and links. \\
\hline
Human inputs and traceability & LaMDA allows user-driven inputs (for example, prompt overrides and configuration choices) and stores run directories, logs, generated artifacts, and parsed summaries for debugging. \\
\hline
\end{tabular}
\end{table}

\begin{table}[t]
\centering
\caption{LLM Roles and Design Flows Autonomy}
\label{Table_S3}
\small
\setlength{\tabcolsep}{5pt}
\renewcommand{\arraystretch}{1.2}
\begin{tabular}{p{0.12\linewidth} p{0.50\linewidth} p{0.32\linewidth}}
\hline
Flow & LLM Roles & Flow Autonomy \\
\hline
Analogue &
Generate Spectre-compliant analogue netlists from user and system prompts for custom design tasks (e.g., OTA/inverter), including parameterised device dimensions and simulation directives, with downstream Spectre parsing/checks. & Prompt-to-netlist generation followed by simulation and metric evaluation. \\
\hline
FPGA &
Generate FPGA design artifacts from prompts (Verilog HDL generation with reasoning on design specifications such as LUT/timing targets). & Large-scale prompt-to-implementation runs across datasets and constraint settings. \\
\hline
RF &
Generate Keysight ADS-compliant RF netlists using prompt context that includes ADS libraries/components, netlist examples, and ADS Python-interface guidance. &
Human-LLM iterative closed-loop refinement (simulation/result feedback across configured iterations). \\
\hline
\end{tabular}
\end{table}

\begin{table}[t]
\centering
\caption{Categorisation of errors across the three design flows}
\label{Table_S4}
\small
\setlength{\tabcolsep}{5pt}
\renewcommand{\arraystretch}{1.2}
\begin{tabular}{p{0.48\linewidth} p{0.44\linewidth}}
\hline
Error Type & Affected Design Entries \\
\hline
Functional behaviour did not match the expected outputs & ResBench: 5, 7, 9, 10, 11, 12, 13, 15, 19, 20, 21, 26, 27, 28, 29, 30, 31, 32, 33, 34, 35, 39, 40, 41, 44, 45, 48, 50, 53, 54, 55 \\
\hline
Procedural assignment targets a signal that should be declared as a register/logic in procedural code & ResBench: 6, 14, 16, 38, 42 \\
\hline
Loop in RTL does not converge for synthesis (likely non-terminating or unsupported loop form) & ResBench: 15, 21, 42 \\
\hline
Source contains invalid or non-printable characters & ResBench: 37 \\
\hline
Too many I/O ports for the target device pins; placement is infeasible & ResBench: 37, 39, 40, 41, 54 \\
\hline
Port/interface mismatch between module definition and instantiation & ResBench: 43 \\
\hline
OTA device-geometry/PDK violations during simulation (generated transistor dimensions outside allowable model ranges for width, length, and area) & OTA \\
\hline
Antenna model/component compatibility failures (undefined ADS models, unknown component classes, missing or invalid substrate/model references, and required-parameter omissions) & Antenna \\
\hline
Antenna ADS netlist structure/parser failures (malformed statements, invalid tokens/parameter syntax, duplicate instance redefinitions, and incorrect connections for components) & Antenna \\
\hline
Antenna simulation-execution environment failures (missing simulation components, local control-file/VM file creation failures, and character-encoding incompatibilities in generated text) & Antenna \\
\hline
\end{tabular}
\end{table}

\begin{table}[t]
\centering
\caption{OTA: System Prompt -- Part I}
\label{Table_S5}
\small
\setlength{\tabcolsep}{5pt}
\renewcommand{\arraystretch}{1.2}
\begin{tabular}{p{0.95\linewidth}}
\hline
You are an expert Cadence Spectre netlist generator for a 5-transistor Operational Transconductance Amplifier (5T OTA).\\
HARD RULES (must follow exactly):\\
- Output ONLY a single fenced code block with language tag "spectre".\\
- Do NOT write explanations or comments before or after the code block.\\
- Use Cadence Spectre syntax (not generic SPICE dot-cards).\\
- Always start with:\\
  simulator lang=spectre\\
  global 0\\
- Use node 0 for ground in all elements.\\
- Do not include any PDK include/model files (they are injected later by a separate script).\\
Target TOPOLOGY (5T OTA):\\
- NMOS differential pair: M1, M2\\
- PMOS diode-connected loads: M3, M4\\
- NMOS tail current source: M5\\
- Single-ended output node: OUT\\
- Differential input nodes: INP, INN\\
- Bias node for tail: VB\\
- Positive supply: VDD\\
- Global ground: 0\\
Parameters (declare exactly these, case-sensitive):\\
- VDD, VCM, VBIAS\\
- WDIFF, WLOAD, WTAIL\\
- LDIFF, LLOAD, LTAIL\\
- CL\\
Sources (use exactly these instance names and node names):\\
- VDD  (VDD 0)   vsource type=dc dc=VDD\\
- VCM  (VCM 0)   vsource type=dc dc=VCM\\
- VBIAS(VB  0)   vsource type=dc dc=VBIAS\\
Small-signal AC excitation:\\
- Use a differential AC source between INP and INN:\\
  VINP  (INP VCM) vsource type=sine mag=1 freq=10\\
  VINN  (INN VCM) vsource type=dc dc=0\\
No other AC sources.\\
Instances (use generic nmos/pmos; the technology-specific device names are added later):\\
- M1 (net03 INP TAIL 0)   nmos w=WDIFF l=LDIFF\\
- M2 (OUT INN TAIL 0)   nmos w=WDIFF l=LDIFF\\
- M3 (net03 net03 VDD VDD)  pmos w=WLOAD l=LLOAD\\
- M4 (OUT net03 VDD VDD)  pmos w=WLOAD l=LLOAD\\
- M5 (TAIL VB  0   0)   nmos w=WTAIL l=LTAIL\\
- C0 (OUT GND) capacitor c=CL\\
\hline
\end{tabular}
\end{table}

\begin{table}[t]
\centering
\caption{OTA: System Prompt -- Part II}
\label{Table_S6}
\small
\setlength{\tabcolsep}{5pt}
\renewcommand{\arraystretch}{1.2}
\begin{tabular}{p{0.95\linewidth}}
\hline
Notes:\\
- INN must be tied to the common-mode node VC (for now we keep it AC-grounded):\\
  (you can either explicitly short INN to VC using a 0V source, or just define node INN = VC in the netlist)\\
- The only output node is OUT.\\
- Do NOT add any extra resistors or capacitors unless explicitly requested.\\
Analyses (exact order):\\
- First, a DC operating point so the OTA is biased:\\
  dcOp dc write="spectre.dc" maxiters=150 maxsteps=10000 annotate=status\\
  dcOpInfo info what=oppoint where=rawfile\\
- Then an AC sweep for open-loop gain and phase:\\
  ac ac start=1 stop=100M dec=60 annotate=status\\
- NEVER use SPICE-style dot commands such as .ac, .dc, .dcOp, .op, .tran.\\
  Always use Spectre analysis syntax like:\\
    ac ac ...\\
    dcOp dc ...\\
    tran tran ...\\
  with NO leading dot on the line.\\
Save options (exactly these lines, in this order):\\
- saveOptions options save=allpub\\
- save VDD:p VBIAS VINP VINN OUT\\
\\
\hline
\end{tabular}
\end{table}

\begin{table}[t]
\centering
\caption{OTA: System Prompt -- Part III}
\label{Table_S7}
\small
\setlength{\tabcolsep}{5pt}
\renewcommand{\arraystretch}{1.2}
\begin{tabular}{p{0.95\linewidth}}
\hline
Additional constraints:\\
- The OTA must be stable with a well-defined low-frequency gain.\\
- The DC operating point must be physically reasonable for the given parameters.\\
- Do NOT add any other sources, analyses, or parameters beyond what is listed above.\\
- Do NOT rename nodes, devices, or parameters: use exactly the names specified here.\\
- Do NOT use curly braces \{ or \} anywhere in the netlist (no JSON, no vsource \{ ... \} blocks).\\
Design heuristics (to improve DC gain reproducibly):\\
- Default priority is meeting DC gain (e.g., $\geq$ 40 dB) before maximizing UGB.\\
- Prefer increasing channel lengths to raise output resistance (ro) rather than using extreme widths or bias current.\\
- Unless the user explicitly requests minimum length, choose:\\
  - LLOAD $\geq$ 2u and LDIFF $\geq$ 2u by default.\\
  - LTAIL $\geq$ 2u if it does not prevent proper biasing.\\
- Keep widths in a reasonable range so the DC operating point remains around mid-supply at OUT.\\
NETLIST HARD CONSTRAINTS (DO NOT VIOLATE):\\
- The netlist MUST contain a single Parameters statement on ONE line:\\
  parameters VDD=... VCM=... VBIAS=... WDIFF=... WLOAD=... WTAIL=... LDIFF=... LLOAD=... LTAIL=... CL=...\\
  Do NOT use double backslash line continuation. Do NOT split this across multiple lines.\\
- The netlist MUST include the following differential input sources EXACTLY (with these node names and structure):\\
  // Differential input sources\\
  VINP  (INP VCM) vsource type=sine mag=1 freq=10\\
  VINN  (INN VCM) vsource type=dc dc=0\\
- Do NOT rename INP, INN, VCM, VINP, VINN or change their role.
\\
\hline
\end{tabular}
\end{table}

\begin{table}[t]
\centering
\caption{RF Flow: Prompt Format}
\label{Table_S8}
\small
\setlength{\tabcolsep}{5pt}
\renewcommand{\arraystretch}{1.2}
\begin{tabular}{p{0.95\linewidth}}
\hline
You are an expert in Python coding and Keysight ADS. Always be precise on syntax and semantics. You are an expert on Python script using Keysight's ADS Design Environment (DE) Python API. This is information regarding libraries, components and simulators:
\{library\_content\}.
It is expected that you generate a netlist like this:
\{netlist\_example\}.
Do not use new line characters in netlist blocks. Keep the netlist elements on a single line. For FR-4 substrate, use the following: model Sub1 MSUB H=1.6 mm Er=4.4 Mur=1 Cond=5.8e7 Hu=1e+33 mm T=0 mm TanD=0.02 Rough=0 Name=Sub1. For an antenna patch, use MLOC component. Consider manufacturing tolerances and practical implementation aspects in your designs.\\
\\
After each non-final iteration, the user responds with a message of the following structure:\\
Requirements: $ S_{11} \le -10\text{ dB} $ at $ 2.4\text{ GHz} $. Simulation results: \texttt{\{main\_result['summary']\}}, where summary is error description or $ S_{11} $ at $ 2.4\text{ GHz} $ and resonant frequency with its $ S_{11} $. Please fix errors, modify the design, and generate an updated netlist.

\\
\hline
\end{tabular}
\end{table}

\begin{table}[t]
\centering
\caption{RF Flow: ADS-Book Format}
\label{Table_S9}
\small
\setlength{\tabcolsep}{5pt}
\renewcommand{\arraystretch}{1.2}
\begin{tabular}{p{0.28\linewidth} p{0.66\linewidth}}
\hline
ADS-Book Feature & Details \\
\hline
Purpose & Local knowledge bundle used to ground RF netlist generation with ADS-specific context. \\
\hline
Root location & LaMDA-RF/Data/ADS-Book/ \\
\hline
ADS Libraries & Exported from local ADS using the Keysight ADS Python API (\texttt{keysight.ads.de}) as described in \texttt{HOW\_TO\_GET\_LIBRARIES.txt}. \newline This process generates plain text files under \texttt{Data/ADS-Book/libraries/} (one \texttt{<library\_name>.txt} per library) listing available components formatted as: \newline \texttt{Library: <name>} \texttt{- Component: <cell>}. \\
\hline
ADS Reference Netlist & Exported from a local ADS schematic using \texttt{design.generate\_netlist()} as described in \texttt{HOW\_TO\_GET\_NETLIST.txt}. \newline The output is saved to \texttt{Data/ADS-Book/Netlist.txt} and is used as a style and syntax template for generated netlists. \\
\hline
Versioning expectation & ADS-Book outputs are generated from local ADS installation and should be regenerated when ADS version or available libraries change. \\
\hline
\end{tabular}
\end{table}

\begin{table}[t]
\centering
\caption{FPGA Flow: Prompt Format}
\label{Table_S10}
\small
\setlength{\tabcolsep}{5pt}
\renewcommand{\arraystretch}{1.2}
\begin{tabular}{p{0.28\linewidth} p{0.66\linewidth}}
\hline
Prompt Element & Content \\
\hline
System prompt &
You are an expert in FPGA design, Verilog coding, Python coding. Be always precise on syntax and semantics. Follow what asked and do not ask more questions.
\\
\hline
Design instruction prompt (appended to user prompt) &
Design the Verilog code of the provided module, following the reported information. Provide these outputs, using these markers in the response: 1. Design name: start marker '// Start Design name', end marker '// End Design name', 2. Verilog design: start marker '// Start Verilog Design', end marker '// End Verilog Design'. Top module name must be the same as the design name. List of syntax rules to follow: - Inputs and outputs must be declared in the top module. - When using blocks, please follow this style: if (condition) begin // code end else begin // code end.
\\
\hline
Dataset-exported user prompt body &
module: \textless module name\textgreater\newline
Problem: \textless natural-language design task\textgreater\newline
Module header: \textless verilog module signature\textgreater
\\
\hline
Actual user prompt sent to LLM in ResBench design generation &
\textless prompt\_content\_from\_dataset\_file\textgreater\newline\newline
\textless DESIGN\_PROMPT\textgreater
\\
\hline
\end{tabular}
\end{table}

\begin{table}[t]
\centering
\caption{RF Antenna Experiments: Iterations To Target}
\label{Table_S11}
\begin{tabular}{llccccc}
\hline
Model & Setting & Attempt & Selected Iter. & Objective (dB) & Result (dB) & \%Var \\
\hline
gpt-4o-mini & Unconstrained & 1 & 1 & -10.0 & -12.7 & 26.6 \\
gpt-4o-mini & Unconstrained & 2 & 2 & -10.0 & -11.9 & 18.6 \\
gpt-4o-mini & Unconstrained & 5 & 4 & -10.0 & -10.5 & 5.5 \\
gpt-4o-mini & Constrained & 3 & 1 & -10.0 & -10.5 & 5.0 \\
gpt-4o-mini & Constrained & 4 & 4 & -10.0 & -12.7 & 26.5 \\
gpt-4o-mini & Constrained & 5 & 4 & -10.0 & -12.6 & 25.6 \\
o1 & Constrained & 1 & 10 & -10.0 & -11.3 & 13.0 \\
\hline
\end{tabular}
\end{table}

\begin{table*}[t]
\centering
\caption{OTA: Ablation Study}
\label{Table_S12}
\begin{tabular}{lllrccc}
\hline
Case & Setup & Model & Mean LLM time (s) & Mean Token Cnt. & Sim. & Obj. \\
\hline
C0 & SP+ / TB+ / C & codestral-2508 & 2.52 & 374.6 & 4/5 & 3/5 \\
C0 & SP+ / TB+ / C & deepseek-chat-v3.1 & 7.03 & 129.6 & 5/5 & 5/5 \\
C0 & SP+ / TB+ / C & gpt-4o & 5.34 & 349.4 & 5/5 & 5/5 \\
C0 & SP+ / TB+ / C & gpt-4o-mini & 10.00 & 359.8 & 5/5 & 5/5 \\
C0 & SP+ / TB+ / C & o1 & 15.95 & 1216.0 & 5/5 & 5/5 \\
C1 & SP- / TB+ / C & codestral-2508 & 13.76 & 1374.2 & 0/5 & 0/5 \\
C1 & SP- / TB+ / C & deepseek-chat-v3.1 & 17.33 & 635.0 & 0/5 & 0/5 \\
C1 & SP- / TB+ / C & gpt-4o & 16.33 & 733.0 & 0/5 & 0/5 \\
C1 & SP- / TB+ / C & gpt-4o-mini & 6.37 & 857.4 & 0/5 & 0/5 \\
C1 & SP- / TB+ / C & o1 & 19.88 & 738.8 & 0/5 & 0/5 \\
C2 & SP+ / TB- / C & codestral-2508 & 6.17 & 380.2 & 0/5 & 0/5 \\
C2 & SP+ / TB- / C & deepseek-chat-v3.1 & 22.45 & 232.8 & 0/5 & 0/5 \\
C2 & SP+ / TB- / C & gpt-4o & 4.59 & 340.6 & 0/5 & 0/5 \\
C2 & SP+ / TB- / C & gpt-4o-mini & 3.19 & 327.6 & 0/5 & 0/5 \\
C2 & SP+ / TB- / C & o1 & 10.90 & 328.0 & 0/5 & 0/5 \\
C3 & SP- / TB- / C & codestral-2508 & 3.05 & 316.8 & 0/5 & 0/5 \\
C3 & SP- / TB- / C & deepseek-chat-v3.1 & 60.43 & 616.0 & 0/5 & 0/5 \\
C3 & SP- / TB- / C & gpt-4o & 8.75 & 309.6 & 0/5 & 0/5 \\
C3 & SP- / TB- / C & gpt-4o-mini & 1.22 & 177.0 & 0/5 & 0/5 \\
C3 & SP- / TB- / C & o1 & 18.59 & 0.0 & 0/5 & 0/5 \\
C4 & SP+ / TB+ / U & codestral-2508 & 3.17 & 377.2 & 1/5 & 0/5 \\
C4 & SP+ / TB+ / U & deepseek-chat-v3.1 & 56.48 & 225.6 & 5/5 & 0/5 \\
C4 & SP+ / TB+ / U & gpt-4o & 4.75 & 346.8 & 5/5 & 4/5 \\
C4 & SP+ / TB+ / U & gpt-4o-mini & 4.85 & 323.2 & 0/5 & 0/5 \\
C4 & SP+ / TB+ / U & o1 & 16.21 & 330.2 & 5/5 & 5/5 \\
C5 & SP- / TB+ / U & codestral-2508 & 9.58 & 899.2 & 0/5 & 0/5 \\
C5 & SP- / TB+ / U & deepseek-chat-v3.1 & 77.87 & 846.8 & 0/5 & 0/5 \\
C5 & SP- / TB+ / U & gpt-4o & 11.01 & 703.6 & 0/5 & 0/5 \\
C5 & SP- / TB+ / U & gpt-4o-mini & 5.55 & 763.0 & 0/5 & 0/5 \\
C5 & SP- / TB+ / U & o1 & 18.41 & 1212.0 & 0/5 & 0/5 \\
C6 & SP+ / TB- / U & codestral-2508 & 5.14 & 386.4 & 0/5 & 0/5 \\
C6 & SP+ / TB- / U & deepseek-chat-v3.1 & 18.67 & 261.6 & 0/5 & 0/5 \\
C6 & SP+ / TB- / U & gpt-4o & 3.69 & 336.0 & 0/5 & 0/5 \\
C6 & SP+ / TB- / U & gpt-4o-mini & 2.96 & 321.8 & 0/5 & 0/5 \\
C6 & SP+ / TB- / U & o1 & 9.74 & 328.0 & 0/5 & 0/5 \\
C7 & SP- / TB- / U & codestral-2508 & 3.98 & 390.0 & 0/5 & 0/5 \\
C7 & SP- / TB- / U & deepseek-chat-v3.1 & 48.63 & 393.6 & 0/5 & 0/5 \\
C7 & SP- / TB- / U & gpt-4o & 3.67 & 262.0 & 0/5 & 0/5 \\
C7 & SP- / TB- / U & gpt-4o-mini & 2.23 & 294.8 & 0/5 & 0/5 \\
C7 & SP- / TB- / U & o1 & 0.00 & 0.0 & 0/5 & 0/5 \\
\hline
\multicolumn{7}{@{}l@{}}{\footnotesize SP: System Prompt; TB: Technology Binder; C/U: Constrained/Unconstrained; +/-: ON/OFF.} \\
\end{tabular}
\end{table*}

\begin{table*}[t]
\centering
\caption{RF Antenna: Ablation Study}
\label{Table_S13}
{\setlength{\tabcolsep}{4pt}
\renewcommand{\arraystretch}{1.1}
\begin{tabular}{lllrcccc}
\hline
Case & Setup & Model & Iters & Mean LLM time (s) & Mean Token Cnt. & Sim. & Obj. \\
\hline
C0 & SP- / ADS- / U & codestral-2508 & 1 & 29.17 & 855.6 & 0/5 & 0/5 \\
C0 & SP- / ADS- / U & deepseek-chat-v3.1 & 1 & 11.96 & 765.0 & 0/5 & 0/5 \\
C0 & SP- / ADS- / U & gpt-4o & 1 & 16.50 & 705.4 & 0/5 & 0/5 \\
C0 & SP- / ADS- / U & gpt-4o-mini & 1 & 5.34 & 713.6 & 0/5 & 0/5 \\
C0 & SP- / ADS- / U & o1 & 1 & 23.60 & 2511.0 & 0/5 & 0/5 \\
C1 & SP- / ADS- / C & codestral-2508 & 1 & 15.95 & 921.0 & 0/5 & 0/5 \\
C1 & SP- / ADS- / C & deepseek-chat-v3.1 & 1 & 27.49 & 830.4 & 0/5 & 0/5 \\
C1 & SP- / ADS- / C & gpt-4o & 1 & 14.13 & 741.0 & 0/5 & 0/5 \\
C1 & SP- / ADS- / C & gpt-4o-mini & 1 & 8.35 & 800.0 & 0/5 & 0/5 \\
C1 & SP- / ADS- / C & o1 & 1 & 27.22 & 3211.2 & 0/5 & 0/5 \\
C2 & SP- / ADS+ / U & codestral-2508 & 1 & 17.92 & 772.6 & 0/5 & 0/5 \\
C2 & SP- / ADS+ / U & deepseek-chat-v3.1 & 1 & 21.41 & 896.2 & 0/5 & 0/5 \\
C2 & SP- / ADS+ / U & gpt-4o & 1 & 12.81 & 700.8 & 0/5 & 0/5 \\
C2 & SP- / ADS+ / U & gpt-4o-mini & 1 & 6.28 & 724.2 & 0/5 & 0/5 \\
C2 & SP- / ADS+ / U & o1 & 1 & 30.79 & 2842.8 & 0/5 & 0/5 \\
C3 & SP- / ADS+ / C & codestral-2508 & 1 & 27.94 & 888.4 & 0/5 & 0/5 \\
C3 & SP- / ADS+ / C & deepseek-chat-v3.1 & 1 & 53.17 & 932.2 & 0/5 & 0/5 \\
C3 & SP- / ADS+ / C & gpt-4o & 1 & 13.06 & 718.0 & 0/5 & 0/5 \\
C3 & SP- / ADS+ / C & gpt-4o-mini & 1 & 8.74 & 830.6 & 0/5 & 0/5 \\
C3 & SP- / ADS+ / C & o1 & 1 & 28.84 & 3486.2 & 0/5 & 0/5 \\
C4 & SP+ / ADS- / U & codestral-2508 & 1 & 28.75 & 781.6 & 0/5 & 0/5 \\
C4 & SP+ / ADS- / U & deepseek-chat-v3.1 & 1 & 27.45 & 832.0 & 0/5 & 0/5 \\
C4 & SP+ / ADS- / U & gpt-4o & 1 & 5.93 & 571.8 & 0/5 & 0/5 \\
C4 & SP+ / ADS- / U & gpt-4o-mini & 1 & 3.79 & 448.6 & 0/5 & 0/5 \\
C4 & SP+ / ADS- / U & o1 & 1 & 18.63 & 2811.2 & 0/5 & 0/5 \\
C5 & SP+ / ADS- / C & codestral-2508 & 1 & 28.06 & 1006.4 & 0/5 & 0/5 \\
C5 & SP+ / ADS- / C & deepseek-chat-v3.1 & 1 & 40.30 & 825.4 & 0/5 & 0/5 \\
C5 & SP+ / ADS- / C & gpt-4o & 1 & 6.77 & 592.4 & 0/5 & 0/5 \\
C5 & SP+ / ADS- / C & gpt-4o-mini & 1 & 3.86 & 496.0 & 0/5 & 0/5 \\
C5 & SP+ / ADS- / C & o1 & 1 & 22.31 & 3124.0 & 0/5 & 0/5 \\
C6 & SP+ / ADS+ / U & codestral-2508 & 10 & 10.43 & 1292.8 & 0/5 & 0/5 \\
C6 & SP+ / ADS+ / U & deepseek-chat-v3.1 & 10 & 44.40 & 901.2 & 0/5 & 0/5 \\
C6 & SP+ / ADS+ / U & gpt-4o & 10 & 7.99 & 619.0 & 2/5 & 0/5 \\
C6 & SP+ / ADS+ / U & gpt-4o-mini & 10 & 9.26 & 801.6 & 4/5 & 3/5 \\
C6 & SP+ / ADS+ / U & o1 & 10 & 25.29 & 1490.8 & 0/5 & 0/5 \\
C7 & SP+ / ADS+ / C & codestral-2508 & 10 & 9.90 & 1098.2 & 0/5 & 0/5 \\
C7 & SP+ / ADS+ / C & deepseek-chat-v3.1 & 10 & 12.77 & 573.0 & 0/5 & 0/5 \\
C7 & SP+ / ADS+ / C & gpt-4o & 10 & 8.62 & 659.8 & 2/5 & 0/5 \\
C7 & SP+ / ADS+ / C & gpt-4o-mini & 10 & 15.96 & 933.2 & 5/5 & 3/5 \\
C7 & SP+ / ADS+ / C & o1 & 10 & 11.57 & 1278.2 & 1/5 & 1/5 \\
\hline
\end{tabular}
}
\\
\parbox{\linewidth}
{\raggedright\footnotesize SP: System Prompt; ADS: ADS tool specifications; C/U: Constrained/Unconstrained; +/-: ON/OFF.\\When Iters=1, the LLM feedback is not included.}
\end{table*}

\begin{table*}[t]
\centering
\normalsize
\caption{FSM: Ablation Study}
\label{Table_S14}
{\setlength{\tabcolsep}{4pt}
\renewcommand{\arraystretch}{1.1}
\begin{tabular}{lllrcccccc}
\hline
Case & Setup & Model & LLM time (s) & Tokens & Verif. & Synth. & Impl. & LUT & Delay \\
\hline
C0 & SP+ / DP+ / C & codestral-2508 & 2.59 & 199.4 & 2/5 & 2/5 & 2/5 & 2/5 & 2/5 \\
C0 & SP+ / DP+ / C & deepseek-chat-v3.1 & 8.72 & 156.0 & 5/5 & 5/5 & 5/5 & 2/5 & 2/5 \\
C0 & SP+ / DP+ / C & gpt-4o & 2.09 & 139.8 & 5/5 & 5/5 & 5/5 & 5/5 & 5/5 \\
C0 & SP+ / DP+ / C & gpt-4o-mini & 4.49 & 150.2 & 5/5 & 5/5 & 5/5 & 5/5 & 5/5 \\
C0 & SP+ / DP+ / C & o1 & 21.29 & 107.2 & 3/5 & 3/5 & 3/5 & 3/5 & 3/5 \\
C1 & SP+ / DP+ / U & codestral-2508 & 2.04 & 208.0 & 0/5 & 0/5 & 0/5 & 0/5 & 0/5 \\
C1 & SP+ / DP+ / U & deepseek-chat-v3.1 & 6.43 & 92.6 & 5/5 & 5/5 & 5/5 & 3/5 & 0/5 \\
C1 & SP+ / DP+ / U & gpt-4o & 2.82 & 142.2 & 5/5 & 5/5 & 5/5 & 5/5 & 0/5 \\
C1 & SP+ / DP+ / U & gpt-4o-mini & 3.48 & 151.0 & 0/5 & 0/5 & 0/5 & 0/5 & 0/5 \\
C1 & SP+ / DP+ / U & o1 & 8.85 & 161.4 & 5/5 & 5/5 & 5/5 & 5/5 & 0/5 \\
C2 & SP- / DP+ / C & codestral-2508 & 3.99 & 349.2 & 1/5 & 1/5 & 1/5 & 1/5 & 1/5 \\
C2 & SP- / DP+ / C & deepseek-chat-v3.1 & 8.68 & 125.4 & 5/5 & 5/5 & 5/5 & 1/5 & 1/5 \\
C2 & SP- / DP+ / C & gpt-4o & 1.95 & 135.2 & 5/5 & 5/5 & 5/5 & 5/5 & 5/5 \\
C2 & SP- / DP+ / C & gpt-4o-mini & 2.07 & 156.0 & 0/5 & 0/5 & 0/5 & 0/5 & 0/5 \\
C2 & SP- / DP+ / C & o1 & 35.74 & 143.2 & 3/5 & 3/5 & 3/5 & 3/5 & 3/5 \\
C3 & SP- / DP+ / U & codestral-2508 & 4.02 & 356.4 & 0/5 & 0/5 & 0/5 & 0/5 & 0/5 \\
C3 & SP- / DP+ / U & deepseek-chat-v3.1 & 13.12 & 144.6 & 5/5 & 5/5 & 5/5 & 5/5 & 0/5 \\
C3 & SP- / DP+ / U & gpt-4o & 2.75 & 163.2 & 5/5 & 5/5 & 5/5 & 5/5 & 0/5 \\
C3 & SP- / DP+ / U & gpt-4o-mini & 4.30 & 159.6 & 0/5 & 0/5 & 0/5 & 0/5 & 0/5 \\
C3 & SP- / DP+ / U & o1 & 9.56 & 153.6 & 5/5 & 5/5 & 5/5 & 4/5 & 0/5 \\
C4 & SP+ / DP- / C & codestral-2508 & 4.39 & 413.6 & 5/5 & 5/5 & 5/5 & 5/5 & 5/5 \\
C4 & SP+ / DP- / C & deepseek-chat-v3.1 & 12.41 & 505.2 & 1/5 & 1/5 & 1/5 & 0/5 & 0/5 \\
C4 & SP+ / DP- / C & gpt-4o & 4.88 & 368.8 & 5/5 & 5/5 & 5/5 & 5/5 & 5/5 \\
C4 & SP+ / DP- / C & gpt-4o-mini & 4.36 & 249.2 & 4/5 & 4/5 & 4/5 & 4/5 & 4/5 \\
C4 & SP+ / DP- / C & o1 & 28.75 & 105.2 & 0/5 & 0/5 & 0/5 & 0/5 & 0/5 \\
C5 & SP+ / DP- / U & codestral-2508 & 4.80 & 480.4 & 0/5 & 0/5 & 0/5 & 0/5 & 0/5 \\
C5 & SP+ / DP- / U & deepseek-chat-v3.1 & 9.36 & 159.0 & 5/5 & 5/5 & 5/5 & 5/5 & 0/5 \\
C5 & SP+ / DP- / U & gpt-4o & 4.72 & 315.6 & 5/5 & 5/5 & 5/5 & 5/5 & 0/5 \\
C5 & SP+ / DP- / U & gpt-4o-mini & 3.74 & 153.2 & 5/5 & 5/5 & 5/5 & 5/5 & 0/5 \\
C5 & SP+ / DP- / U & o1 & 8.26 & 110.8 & 5/5 & 5/5 & 5/5 & 5/5 & 0/5 \\
C6 & SP- / DP- / C & codestral-2508 & 5.55 & 461.0 & 4/5 & 4/5 & 4/5 & 4/5 & 4/5 \\
C6 & SP- / DP- / C & deepseek-chat-v3.1 & 66.92 & 1321.2 & 1/5 & 1/5 & 1/5 & 0/5 & 1/5 \\
C6 & SP- / DP- / C & gpt-4o & 5.02 & 439.8 & 4/5 & 4/5 & 4/5 & 4/5 & 4/5 \\
C6 & SP- / DP- / C & gpt-4o-mini & 9.21 & 502.2 & 3/5 & 3/5 & 3/5 & 3/5 & 3/5 \\
C6 & SP- / DP- / C & o1 & 38.59 & 165.2 & 0/5 & 0/5 & 0/5 & 0/5 & 0/5 \\
C7 & SP- / DP- / U & codestral-2508 & 4.59 & 375.4 & 2/5 & 2/5 & 2/5 & 2/5 & 0/5 \\
C7 & SP- / DP- / U & deepseek-chat-v3.1 & 16.36 & 367.8 & 2/5 & 2/5 & 2/5 & 2/5 & 0/5 \\
C7 & SP- / DP- / U & gpt-4o & 5.63 & 461.0 & 4/5 & 4/5 & 4/5 & 4/5 & 0/5 \\
C7 & SP- / DP- / U & gpt-4o-mini & 5.94 & 427.8 & 5/5 & 5/5 & 5/5 & 5/5 & 0/5 \\
C7 & SP- / DP- / U & o1 & 8.01 & 183.6 & 5/5 & 5/5 & 5/5 & 5/5 & 0/5 \\
\hline
\end{tabular}
}
\\
\parbox{\linewidth}{\raggedright\footnotesize SP: System Prompt; DP: Design Prompt; C/U: Constrained/Unconstrained; +/-: ON/OFF. \\LLM time and token count are mean values.}
\end{table*}

\end{document}